\documentclass[namedreferences]{solarphysics}
\usepackage[hyperref,optionalrh,solaromanenum]{spr-sola-addons} 
\usepackage{graphicx}                    
\usepackage{relsize}
\usepackage{txfonts}

\usepackage{amsmath}
\usepackage{amssymb}                    
\usepackage{color}                       
\usepackage{breakurl}                         
\usepackage{lineno}

\begin{document}

\begin{article}

\begin{opening}


\title{The Magnetic Field Geometry of Small Solar Wind Flux Ropes Inferred from their Twist Distribution}

%
\author[addressref={aff1},corref,email={unhyuan@gmail.com}]{\inits{W. Y.}\fnm{W. Yu}\lnm{}}  
\author[addressref={aff1},corref]{\inits{C.J. F.}\fnm{C.J. Farrugia}\lnm{}}
\author[addressref={aff1},corref]{\inits{N. L.}\fnm{N. Lugaz}\lnm{}}
\author[addressref={aff1},corref]{\inits{A.B. G.}\fnm{A. B. Galvin}\lnm{}}
\author[addressref={aff2},corref]{\inits{C. M.}\fnm{C. M\"ostl}\lnm{}}
\author[addressref={aff1},corref]{\inits{K. P.}\fnm{K. Paulson}\lnm{}}
\author[addressref={aff3},corref]{\inits{P. V.}\fnm{P. Vemareddy}\lnm{}}

%
\runningauthor{Yu {\it et al.}}
\runningtitle{Appraising Force-Free Models for Small Flux Ropes}

\address[id={aff1}]{Space Science Center and Department of Physics, University of New Hampshire, Durham, New Hampshire, USA. }
\address[id={aff2}]{Space Research Institute, Austrian Academy of Sciences, Graz, Austria}
\address[id={aff3}]{Indian Institute of Astrophysics, Koramangala, Bengaluru-560034, India}


\begin{abstract}

This work extends recent efforts on the force-free modeling of large flux rope-type structures (magnetic clouds, MCs)
to much smaller spatial scales. We first select small flux ropes (SFRs) by eye whose duration is unambiguous and which were observed by the {\it Solar Terrestrial Relations Observatory} (STEREO) or {\it Wind} spacecraft during solar maximum years. We inquire into which analytical technique is physically most appropriate, augmenting the numerical modeling with considerations of magnetic twist.  The observational fact that these SFRs typically do not expand significantly into the solar wind makes static models appropriate for this study.  SFRs can be modeled with force-free methods during the maximum phase of solar activity. We consider three models: (i) linear force-free field ($\bigtriangledown\times$ {\bf B} = $\alpha ({\bf r})$ {\bf B}) with a specific, prescribed constant $\alpha$ (Lundquist solution), and (ii) with $\alpha$ as a free constant parameter (``Lundquist-alpha'' solution), (iii) uniform twist field (Gold-Hoyle solution). We retain only those cases where the impact parameter is less than one-half the FR radius, $R$, so the results should be robust (29 cases). The SFR radii lie in the range [$\sim$ 0.003, 0.059] AU. Comparing results, we find that the Lundquist-alpha and uniform twist solutions yielded comparable and small normalized $\chi^2$ values in most cases. So analytical modeling alone cannot distinguish which of these two is better in reproducing their magnetic field geometry. We then use Grad-Shafranov (GS) reconstruction to analyze these events further in a model-independent way. The orientations derived from GS are close to those obtained from the uniform twist field model. We then considered the twist {\it per} unit length, $\tau$, both its profile through the FR and its absolute value, applying a graphic approach to obtain $\tau$ from the GS solution. The results are in better agreement with the uniform twist model. We find $\tau$ to lie in the range [5.6, 34] turns/AU, {\it i.e.},  much higher than typical values for MCs. The GH model-derived $\tau$ values are  comparable to those obtained from GS reconstruction. We find that twist {\it per} unit length ($L$) is inversely proportional to $R$, as $\tau \sim 0.17/R$.
We combine MC and SFR results on $\tau (R)$ and give a relation which is  approximately valid for both sets.
Using this, we find that the axial and azimuthal fluxes, $F_z$ and $F_\phi$, vary as 
$\approx 2.1 B_0 R^2$ ($\times10^{21}$ {\textrm Mx}) and $F_{\phi}/L \approx 0.36 B_0 R$ ($\times10^{21}$ {\textrm Mx/AU}). 
The relative helicity {\it per} unit length, $H/L \approx 0.75 B_0^2 R^3$ ($\times10^{42}$ {\textrm Mx$^2$/AU}).

\end{abstract}


\end{opening}

\section{Introduction}

Small magnetic flux ropes (SFRs), with a duration of 12 hours or less, are regions characterized by an enhanced magnetic field strength, a large rotation of the magnetic field vector, and sometimes additional plasma characteristics.  Specifically, the plasma characteristics used to define SFRs are low proton $\beta$ (ratio of the proton pressure to the magnetic pressure) or low proton temperature, and low Alfv\'en Mach number \citep{Yu2016}.
These properties make these structures similar to magnetic clouds (MCs), {\it i.e.}, large-scale structures initially described by \inlinecite{Burlaga1981}. SFRs have been identified in the solar wind for over two decades  now \citep{Moldwin2000}. Recent research have focused on three main topics: (i) the nature and origin of these SFRs: coronal or heliospheric \citep{Moldwin2000}, (ii) their relation to large flux ropes, {\it i.e.}, MCs \citep{Janvier2013}, and (iii) their relation to turbulence and particle acceleration \citep{Hu2015}. In order to address these outstanding questions, it is necessary to determine the physical properties of these SFRs, such as their orientation, radius, current density, magnetic field line twist, relative helicity, toroidal (axial) and poloidal (azimuthal) magnetic fluxes, among others. In order to do so, modeling, least-squares fitting and reconstruction techniques are required. It is well understood that the choice of the technique can give highly varying results, especially for the flux rope orientation \citep{Riley2004, Al-Haddad2013, Al-Haddad2018, Wang2016} but also for the field line twist, defined by $\tau(r) = B_{\phi}/(rB_z)$ in local cylindrical coordinates. 
This quantity is strongly model-dependent  \citep{Dasso2006}. 
The goal of this work is twofold: (1) to determine how the choice of model affects the derived characteristics of SFRs, and which technique is {\em physically} most appropriate for these small transients; and (2) to determine how flux rope properties, especially the twist (number of turns {\it per} {\textrm AU}), change as one considers smaller flux ropes, much smaller than typical MC sizes, whose average radius is $\sim$0.15 {\textrm AU}. These two issues are intimately linked as different models yield different twist distributions.

When modeling MCs, 
a linear force-free solution ({\it i.e.} Lundquist, 1950) is still the most prevalent model, following the initial idea of \inlinecite{Burlaga1988}. Data fitting to this model was implemented by \inlinecite{Lepping1990, Lepping2006}, among others. In this model the force-free equation, $\bigtriangledown \times {\bf B} = \alpha({\bf r}) {\bf B}$, is solved with $\alpha$ = 2.404/R so that the axial
field vanishes at the periphery of the MC (at $r = R$). In order to perform least-squares fitting of the data to this model, it is necessary to identify the boundaries of these flux ropes.
Improvements and modifications have been proposed, including a curved axis \citep{Marubashi2007} and self-similar expansion
({\it e.g.} Farrugia {\it et al.}, 1993; Vemareddy {\it et al.}, 2016). One of the main properties of the Lundquist linear force-free model is that the twist in the flux rope is a minimum on the axis (= $\alpha$/2) and increases without bound at the FR periphery.

There are only few observations that can constrain this model assumption.  
In an interesting study,  \inlinecite{Larson1997}  derived the field-line lengths inside one magnetic cloud by studying the propagation times of solar energetic electrons.  They chose a case where an impulsive energetic solar electron event was associated with a type III radio burst. By timing the arrival of the electrons as a function of energy, they could estimate of lengths of the magnetic field lines. The result was qualitatively in accord with the
Lundquist  model in that the field line length increased toward  the periphery of the MC.
Solar models of flux ropes generated by reconnection in association with solar flares yielded structures which were typically less twisted near the axis, and more twisted to the boundaries (see  {\it e.g.},  vanBallegooijen and Mackay, 2007; Aulanier, Janvier, and Schmieder, 2012). This, too, is consistent  with the Lundquist's model used for {\it in situ} measurements.

\inlinecite{Kahler2011} extended Larson's {\it et al.} (1997) study by discussing more events but using the same approach. The authors derived a range of lengths 1.3 to 3.7 AU for MCs.  However, 
their results exhibited  poor correlation between the computed electron path lengths and the Lundquist's model magnetic field line
lengths: the expected long path lengths near the MC boundary were absent. This conclusion is at variance with the Lundquist model.

A number of studies have found evidence that large flux ropes associated with MCs or interplanetary coronal mass ejections (ICMEs) may be characterized by a nearly-uniform twist throughout the flux rope.  A first example of a uniform twist flux rope in the interplanetary medium was given by \inlinecite{Farrugia1999}.
Using Grad-Shafranov reconstruction, \inlinecite{HS2002} examined the MC on 18 October 1995  and came to a surprising conclusion: when they evaluated the twist, $\tau$, it appeared not to increase with increasing distance from the axis.
Another example is the work of \inlinecite{Liu2008} and \inlinecite{Moestl2009} who studied an MC observed  simultaneously by the {\it Solar Terrestrial Relations Observatory} (STEREO) and {\it Wind} spacecraft, with a separation of order 0.1 AU. They showed how data at one spacecraft can be used to predict the observations at another spacecraft distant from the first by about the radius of the MC.
In addition, \inlinecite{Moestl2009} used the graphic approach (their Figure 2; see further below) to the solution of the Grad-Shafranov (GS) equation \citep{Hu2001} to obtain the field line twist distribution. It varied between 1.5 and 1.7 turns/AU, {\it i.e.} the twist distribution was found not to change much with distance from the axis. This was confirmed by other studies using the same model \citep{Hu2015}. As such, the Gold-Hoyle solution (GH: Gold and Hoyle, 1960), in which the field line twist is constant, may be better suited than the Lundquist solution \citep{Lundquist1950} to reproduce the field line geometry.

There were also studies which included sets of MCs.  Thus \inlinecite{Hu2014} analyzed 18 magnetic flux ropes and they found that half of them had a relatively low twist distribution with values inside the MCs ranging from 1.7 and 7.7 turns/AU. The events with high twist were found to show a decrease of the twist away from the axis. \inlinecite{Hu2015} studied the field line twist and length distributions within magnetic flux ropes using the GS-reconstruction. Their result suggested that MC field lines were uniformly twisted. They suggested that the GH solution, rather than the Lundquist solution, should be used to model  large magnetic flux ropes. 
In \inlinecite{Wang2016}, the authors applied a velocity-modified uniform twist force-free flux rope model to 115 MCs. 
Considering only their quality 1 events (52 cases), they obtained radii lying between 0.02 and 0.3 AU, and twists between 1 and 14 turns/AU. While their results, too, support the GH model, they found that the model-derived twists were probably overestimated by a factor of 2.5  compared with those derived from energetic electron events and by 1.4 compared to
results from GS reconstruction. Vemareddy {\it et al.} (2016) compared properties of an MC with its solar source. Analyzing the interplanetary measurements, they found that the field line twist of the MC decreased from the axis toward the edges.

In this article, the question of the magnetic field geometry of flux ropes as seen through the distribution of their field line twist will be extended to smaller scales. In previous work, we presented a study of small solar wind transients of duration less than 12 hours \citep{Yu2014}, a subset of which were SFRs.
In a subsequent, large study spanning several years of observations from STEREO and {\it Wind} \citep{Yu2016}, we showed statistical results for the $\beta_{plasma}$ of SFRs observed by {\it Wind} from year 1995 to 2014. The value of $\beta_{plasma}$ was found to depend on the solar activity level. During solar minimum years, $\beta_{plasma}$ is close to unity and non force-~free models should be used. However, the $\beta_{plasma}$ is less than 1 in the solar maximum years, and thus force-free models could be appropriate. Here, we propose to build upon this finding and compare force-free models in a dedicated study targeting SFRs near solar maximum.

The layout of the article is as follows. In Section~\ref{s:models}, we present the models used in the rest of the article, namely the GH, the classical Lundquist fitting, in which the parameter $\alpha= 2.404/R$,  and a variant of the Lundquist solution for which $\alpha$ is a constant, fitted parameter. We supplement the fitting results by
GS reconstruction so as to add considerations of field line twist. In Section~\ref{s:methodology} we discuss our criteria for data selection. In Section \ref{s:examples} we then present three case studies of SFRs measured by {\it Wind} or  STEREO and test the different models used in this study.
We then present more general results 
for 29 well-observed SFRs in Section~\ref{s:statistics}. By ``well observed'' we mean that they all constitute visually very clear FRs and, in addition, all models can fit them successfully (with the goodness-of-fit parameter $\chi^2 < 0.2$) and with an impact parameter $<$ 0.5 R.  Statistical results on these small flux ropes (8 SFRs from  STEREO-A (STA), 10 SFRs from  STEREO-B (STB), 11 SFRs from {\it Wind}) for the year 2000--2003 and 2012--2014, the size and twist distribution  and relation between twist and radius are investigated. We discuss the results and draw our conclusions in Section~\ref{s:discussion}.


 \section{Analytical Models}%
\label{s:models} 

We now describe the three analytical models we use. We note that all three are static models so they
are only useful if the SFRs either do not expand at all, or their expansion speed is much less than the bulk speed of the transient. To anticipate, we show below that this is indeed the case.

\subsection{Linear Force-Free Field}

The magnetohydrostatic solution of a force-free magnetic field,  $\bigtriangledown \times {\bf B} = \alpha {\bf B}$, with constant $\alpha$  was obtained by \inlinecite{Lundquist1950} (called Lundquist model in this article). In cylindrical coordinates (r, $\phi$, z) this solution is given by 

\begin{eqnarray}
   \begin{cases}
    B_r = 0
    \\
    B_{\phi} = H B_0 J_1(\alpha r)
    \\
    B_z = B_0 J_0(\alpha r),
  \end{cases}
\end{eqnarray}
where $J_n$ is the Bessel function of the first kind of order n, $B_0$ is the strength of the magnetic field at the axis, and H indicates the handedness of the magnetic field (positive for right-handed and negative for left-handed).  Parameter $\alpha = 2.404/R$ ($R$ = radius of the FR), coinciding with the first zero of $J_0$ on the FR boundary, where the axial field, $B_z$, vanishes.
The magnetic field line twist {\it per} unit length, {\it $\tau(r) = B_{\phi}/(2\pi rB_z)$}, is: 
\begin{equation}
 {\it \tau(r) = HJ_1(\alpha r)/(2\pi rJ_0(\alpha r)).}
\end{equation}
Parameter $\tau$ is in units of turns/AU. $\tau(0) = \alpha/4\pi$
is the twist at the flux rope axis. The quantity $\tau$ increases without bound as we go to the boundary of the FR.

The Lundquist solution with non-fixed $\alpha$ value (``Lundquist-alpha'' solution, abbreviated LA) is one where we do not require the boundary of the SFR to be the first zero of $J_0$. In this case $\alpha$ is obtained from least-squares fitting, with the 
restriction that  $\alpha < 2.404/R$ so that the axial field does not reverse direction inside the FR.

\subsection{Nonlinear Force-Free Field: Uniformly Twisted Field}

The solution of the magnetic field of a uniformly twisted FR was given by \inlinecite{GH1960}:
\begin{eqnarray}
   \begin{cases}
\larger
    B_r = 0
     \\
    B_{\phi} = 2\pi \tau r B_0 / (1+4 \pi^2 \tau^2 r^2)
    \\
    B_z = B_0 / (1+4 \pi^2 \tau^2 r^2),
  \end{cases}
\end{eqnarray}
 where $B_0$ is the strength of the magnetic field at the axis. The quantity $\tau$ is the magnetic field line twist, which is independent of r, and expressed in units of turns/AU.

\subsection{GS-Reconstruction}

The Grad-Shafranov reconstruction technique is applicable to magnetohydrostatic structures possessing an invariant direction ({\it i.e.} axis of FR). 
It was first applied in a magnetopause context by \inlinecite{HS1999} and in a  magnetic cloud context by \inlinecite{Hu2001, HS2002}.
 We give here a brief description. More details can be found in  Hu and Sonnerup (2002), Sonnerup {\it et al.} (2006), M\"ostl {\it et al.} (2009).
The method depends on first transforming the magnetic field and plasma data to a co-moving frame, usually the  deHoffmann-Teller frame where the flow is field-aligned.  This is done by minimizing the convective electric field, -${\bf v} \times {\bf B}$ \citep{Khrabrov1998}. 
In this frame  the condition for magnetohydrostatic equilibrium, {\it i.e.} ${\bf j} \times {\bf B} = -\nabla P$ is given by 
the Grad-Shafranov equation, which expresses a relation between  the vector potential A, the axial field $B_z$, and the sum of the thermal and axial magnetic pressure ({\it i.e.} the transverse pressure $P_t$) and

\begin{equation}
    \frac{\partial^2 A}{\partial x^2} + \frac{\partial^2 A}{\partial y^2}= -\mu_0\frac{d}{dA}(P+\frac{B^2_z}{2\mu_0}) = -\mu_0\frac{dP_t(A)}{dA},
\end{equation}
where {\bf B} is given by
\begin{equation}
{\bf B} = (\frac{\partial A}{\partial y}, \frac{-\partial A}{\partial x}, B_z) .
\end{equation}

Note that it is not required that the structure be force-free.
The pressure $P_t$ is a  function of A only \citep{Sturrock1994}.  Enforcing the single-valuedness of $P_t$ as a function of  A yields,
when successful, the axis orientation of the flux rope and the closest distance the spacecraft passes from the 
FR axis ({\it i.e.} the impact parameter). To check this, a polynomial fit of $P_t$(A) is examined, and the 
fitting residue, $R_f$, gives a measure of how good this single-valued requirement is satisfied (quality of fit).  Note that this might be satisfied within a radius which is less than that inferred from the data.
In general, an $R_f$ not exceeding 0.20 is considered acceptable \citep{Hu2014}.  This, then, gives the axis direction and impact parameter.

The best fit of $P_t$(A) gives the right-hand side of Equation 4.  A Grad-Shafranov solver is employed to solve Equation 4, based on a Taylor expansion of the  solution away from the spacecraft trajectory. The resulting solution is a magnetic field map which is  presented in the transverse XY plane as closed contours of A (see {\it e.g.} Figure 2).
We then apply a graphic approach to the GS solution \citep{Moestl2009, Hu2014, Hu2015, Vemareddy2016}
from which we  obtain the twist distribution in turns/AU. In the graphic method, various field lines at different distances from the axis ({\it i.e.} different A values) are traced until they make one complete turn around the axis. This gives the pitch of the field lines, from which $\tau$ can be derived.

\section{Methodology}
\label{s:methodology}

In this work, we chose 29 SFRs from the lists in \inlinecite{Yu2016}. They all occurred during the solar activity maximum years 
(2000--2003, 2012--2014)
where $\beta_{plasma} \ll 1$. Of these, 8 events were observed by STEREO-A (STA), 10 by STEREO-B (STB), and 11 by the {\it Wind} spacecraft. All these  FRs are fitted by (i) Lundquist solution with fixed alpha (Lundquist solution, L), 
(ii) Lundquist solution with free constant alpha (Lundquist-alpha solution, LA),
and (iii) uniform twist solution (Gold-Hoyle solution, GH). They are then reconstructed using the GS technique. The fitting results are  shown in Table 1 (STA), Table 2 (STB), and Table 3 ({\it Wind}). The twist, $\tau$, for these events is studied in the following manner: for L and LA solutions, we study the average $\tau$ along the trajectory; for GH, we obtain $\tau$ directly from  fitting; and for GS-reconstruction, $\tau$ is derived by applying the  graphic approach, from which we take the average value.

Our fitting procedure is designed to fit the three observed magnetic field components. There are five free parameters (axis orientation ($\theta$, $\phi$), magnetic field strength on the SFR axis ($B_0$), handedness (H), impact parameter) for the L solution, six free parameters ($\theta$, $\phi$, $B_0$, H, impact parameter, $\alpha$) for the LA solution, and five free parameters ($\theta$, $\phi$, $B_0$, impact parameter, twist) for the GH solution.

First, we give initial values for these free parameters. The $\theta$ and $\phi$ are set to be the orientation ($\theta_{min}$ and $\phi_{min}$) obtained from minimum variance analysis of the magnetic field \citep{Sonnerup1998}, $B_0$ is initially set to be 10 nT, H is set to be 1 or -1, the impact parameter is set to be $0.01 \times V_{SW} \times \Delta t$, and the twist is set to be 1. Second, we transform the magnetic field components with the initial values in the SFR coordinate (Equations 1 and 3) to the Geocentric-Solar-Ecliptic (GSE) coordinate system. We then evaluate the goodness of fit by calculating the normalized $\chi^2$ of the difference between the modeled and the observed magnetic field components. 
\begin{equation}
    \chi^2 = \frac{1}{3N} \sum_{i=1}^{N}[(B_{xi}^m - B_{xi}^o)^2 + (B_{yi}^m - B_{yi}^o)^2 + (B_{zi}^m - B_{zi}^o)^2] / (B_{ti}^o)^2, 
\end{equation}
where N is the number of measurements and $B_x, B_y, B_z$ are the three magnetic field components in the GSE coordinate system while $B_t$ is the total magnetic field strength. Third, we use the Image Data Language (IDL) least-squares fitting procedure MPFITFUN \citep{Markwardt2009} to find the best fit. We require $\theta$ to be from $\theta_{min}-30^{\circ}$ to $\theta_{min}+30^{\circ}$, and $\phi$ from $\phi_{min}-30^{\circ}$ to $\phi_{min}+30^{\circ}$.  Then, we require that the impact parameter should be less than 0.5 times the radius of the SFR, and $\chi^2$ less than 0.2. The fitting routine evaluates the user function and parameter limitations and returns the best fit with the smallest value of $\chi^2$.

For events observed by the STEREO spacecraft, the data used are from In-Situ Measurements of Particles (IMPACT: \citep{Luhmann2008}) and Plasma and Suprathermal Ion Composition (PLASTIC:  \citep{Galvin2008}) instrument suites at 1 min resolution. To derive the electron contribution to the plasma pressure we apply the result of \inlinecite{Newbury1998}, who gave an estimate for electron temperature, $T_e$, based on data from the {\it International Sun-Earth Explorer} (ISEE) spacecraft. They obtained $T_e = 1.42 \times 10^5 K$. We use this value for $T_e$ and let $N_e = N_p$, same electron and proton densities, when we calculate the $\beta_{plasma}$. For events observed by the {\it Wind} spacecraft, we use key parameter data acquired by the {\it Magnetic Fields Investigation} (MFI: \citep{Lepping1995}) and {\it Solar Wind Experiment} (SWE: \citep{Ogilvie1995}). The magnetic field data are at 1 min while proton data are at 92 s time resolutions. The electron data are also obtained from the SWE instrument; they have a temporal resolution of 6-12 s before May 2001, and 9 s after August 2002.

We base our identification of the SFR boundaries on a visual inspection of the data to determine the beginning and end of a clear rotation. Often this rotation starts and ends at discontinuities, making it easier to identify its extent. Then we confirm this by doing a minimum variance analysis of the magnetic field data to see if the routine finds a robust flux rope axis orientation from the ratio of intermediate-to-minimum eigenvalues.   The data are not averaged.

\section{Data Examples}
\label{s:examples}

We now discuss three case studies of SFRs observed by STA, STB or {\it Wind}.   They are presented in order of increasing size.

\subsection{Event 1: STA - 6 May 2014}

This event was recorded by STA on 6 May 2014. Figure 1 {\it (left)} shows the observed magnetic field and proton data. 
From top to bottom, the panels show  the total magnetic field and its components in Radial Tangential Normal (RTN) coordinates, the bulk proton speed,
$V_p$, density, $N_p$, and 
temperature, $T$ (in black is the proton temperature $T_p$, in red is the expected temperature $T_{exp}$, and in green is the electron temperature $T_e$),
the Alfv\'en Mach number, $M_A$, and $\beta$ (in black is the $\beta_{proton}$, and in red the $\beta_{plasma}$, which includes both proton and electron contributions).  In the RTN system,  R points from the Sun  to the spacecraft, T is parallel to the solar equatorial plane and points in the direction of planetary motion, and N completes the right-handed RTN triad, which is such that the RN plane contains the rotation axis of the Sun.
Overlaid on the data in the first four panels are
the fitting results: lightblue, red and blue tracers are for the L, LA, and GH models, respectively. 
For clarity, the first four panels and the numerical fits are shown in an expanded form on the right.

\begin{figure}[!ht]
\begin{center} 
\includegraphics[width=4.8in]{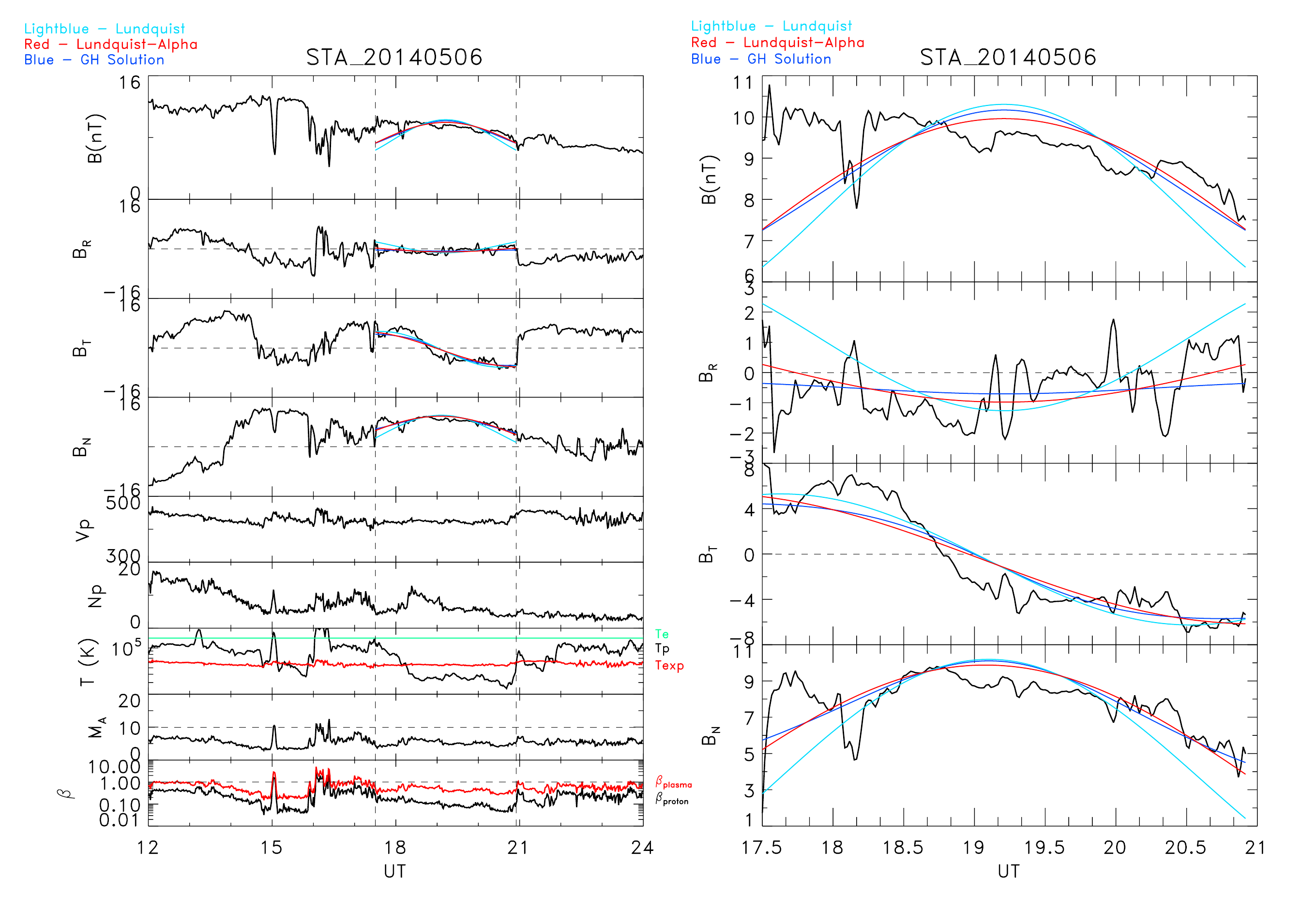}
\end{center} 
 \caption{Event 1: The small flux rope (SFR) observed by STA on 6 May 2014. The SFR interval lies between the vertical lines and has a 
duration of 3.42 hrs. Data are plotted by black traces, while lightblue, red, and blue traces indicate the fitting results 
obtained from the L, LA and GH models, respectively.  For clarity, the plot on the right shows the first four panels in an expanded form.}
\label{fig:event1 - analytical fitting results}
\end{figure}

\begin{figure}[!ht]
\begin{center} 
\includegraphics[width=4.8 in]{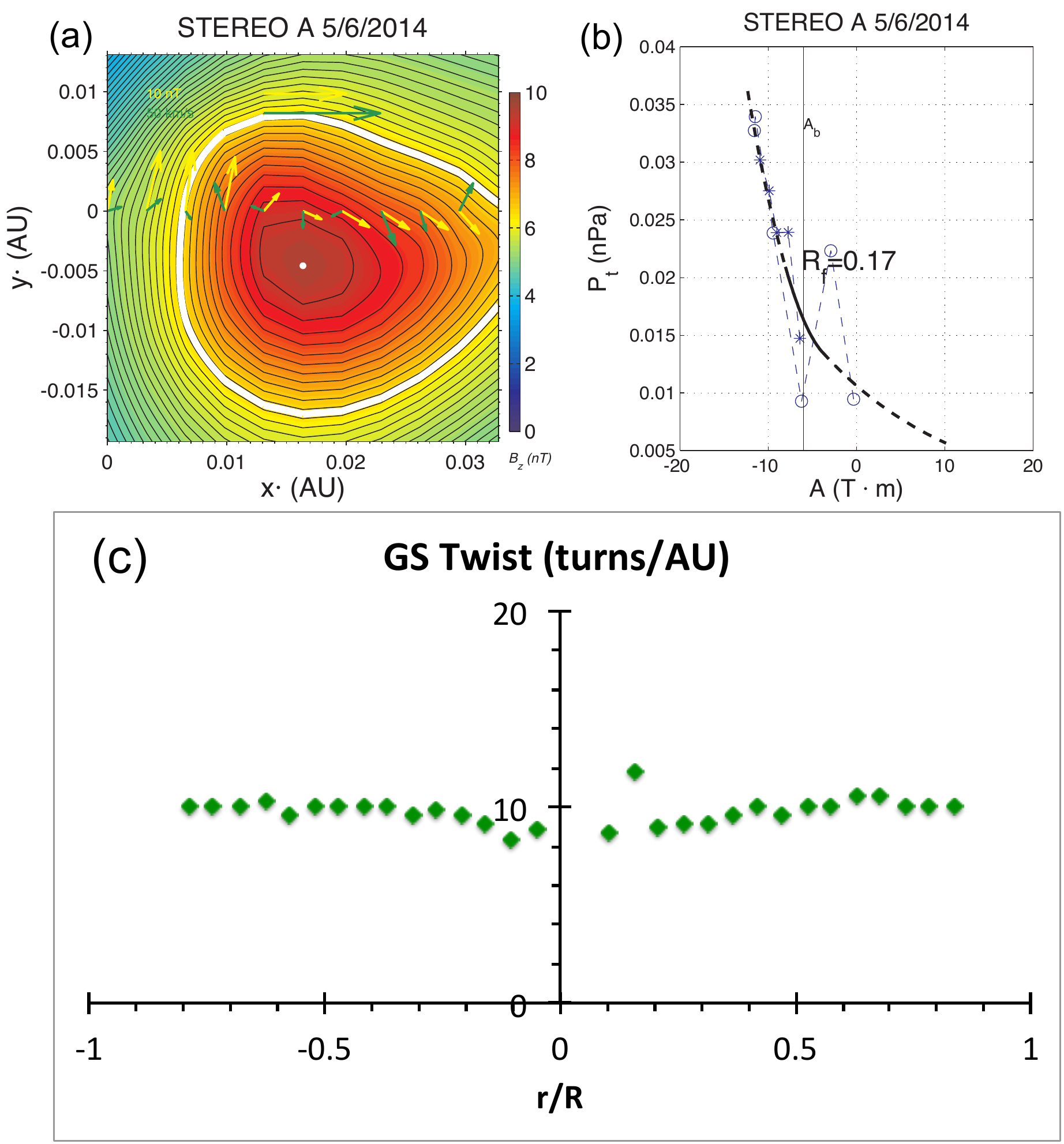}
\end{center} 
 \caption{Event 1: The SFR observed by STA at 6 May 2014. GS-reconstruction results: (a) magnetic field map, (b) fitting residue,
$R_f$, (c) twist distribution from GS-reconstruction. The vertical line in (b) gives the boundary value for $A$ within which the GS reconstruction is most reliable.  It corresponds to the thick white curve in the map of the magnetic field (a).
In panel a the arrows indicate the re-sampled measurements of the 
magnetic field (yellow) and velocity (green) vectors along the inferred spacecraft path. }
\label{fig:event1 - GS results}
\end{figure}

The SFR was observed from 17:30:00 to 20:55:00 UT, giving it a  duration of 3.42 h. It has large and relatively smooth rotations of the magnetic field components. 
$T_p$ is lower than the expected temperature, making it a small MC after Burlaga's {\it et al.} (1981) definition. 
$T_e$ is much higher than $T_p$. $M_A$ is low, of order 5, and $\beta_{proton}$ and $\beta_{plasma}$ are  smaller than 1, so the structure is magnetically dominated. 
The results of the least-squares fitting for  the magnetic field data from the three analytical force-free models are shown in the top four panels 
and Figure 1 (right)
by the colored traces. All the fitting details are given in Table 1, which provides details on SFRs observed by STA and arranged by date.

From the figure and Table 1  we can see that all three models fit the magnetic field components very well. 
The normalized $\chi^2$ values from all models are comparable and small: 0.03, 0.02, and 0.02 for the L, LA and GH models, respectively.  
All three models give H = -1, {\it i.e.} left-hand chirality, see Table 1.
The orientations obtained from the L and LA models are within $21.5^{\circ}$ of each other, while the orientation 
obtained from GH makes an angle of 23 and $1.5^{\circ}$ with the L and LA model, respectively. 
The fitted magnetic field strength at the axis and the radius from these three solutions are close, of order 10.3 nT and 0.017 AU, respectively.  The GH solution in this event has the smallest impact parameter (0.269 $R$).  We conclude that, based on the fitting results alone, it is not possible to choose between the models.

Figure 2 shows the results obtained from the GS-reconstruction. They are: (a) magnetic field map in a plane perpendicular to the axis, where each contour corresponds to a given value of $A$ , (b) fitting residue, $R_f$,  and (c) the twist distribution from GS-reconstruction by using the graphic approach.  The colors in panel a give the strength of the axial field with scale on the right.  The strength of the axial field peaks at the white dot. The yellow and green arrows show a sample of the field and flow vectors on the inferred spacecraft trajectory. The heavy white curve gives the boundary of the FR as determined by the technique.
Further details of the GS-reconstruction are given in Table 1. 
From panel a the magnetic contours show that the cross-section of the FR tends to be circular, 
and the spacecraft passes close to the center of the FR. It has a left-handed chirality, the same as in the three analytical models. The fitting residue in Figure 1b is $\sim$0.17, which is good.  The orientation of this SFR is within $4^{\circ}$ of that obtained from the GH solution.

The twist distribution obtained from the GS-reconstruction is shown in Figure 2c by the green symbols. The two branches correspond to values obtained as the spacecraft moves towards the axis (left) and away from it. Except near the axis, the twist values are fairly constant at $\approx$ 9.8 turns/AU. This is the opposite trend to that of LA model, which diverges as $R$ increases.  
The average twist of the GS model is practically the same as that obtained from GH fitting (10.0 turns/AU).

\subsection{Event 2: STB - 24 April 2013}

\begin{figure}[!ht]
\begin{center} 
\includegraphics[width=4.8 in]{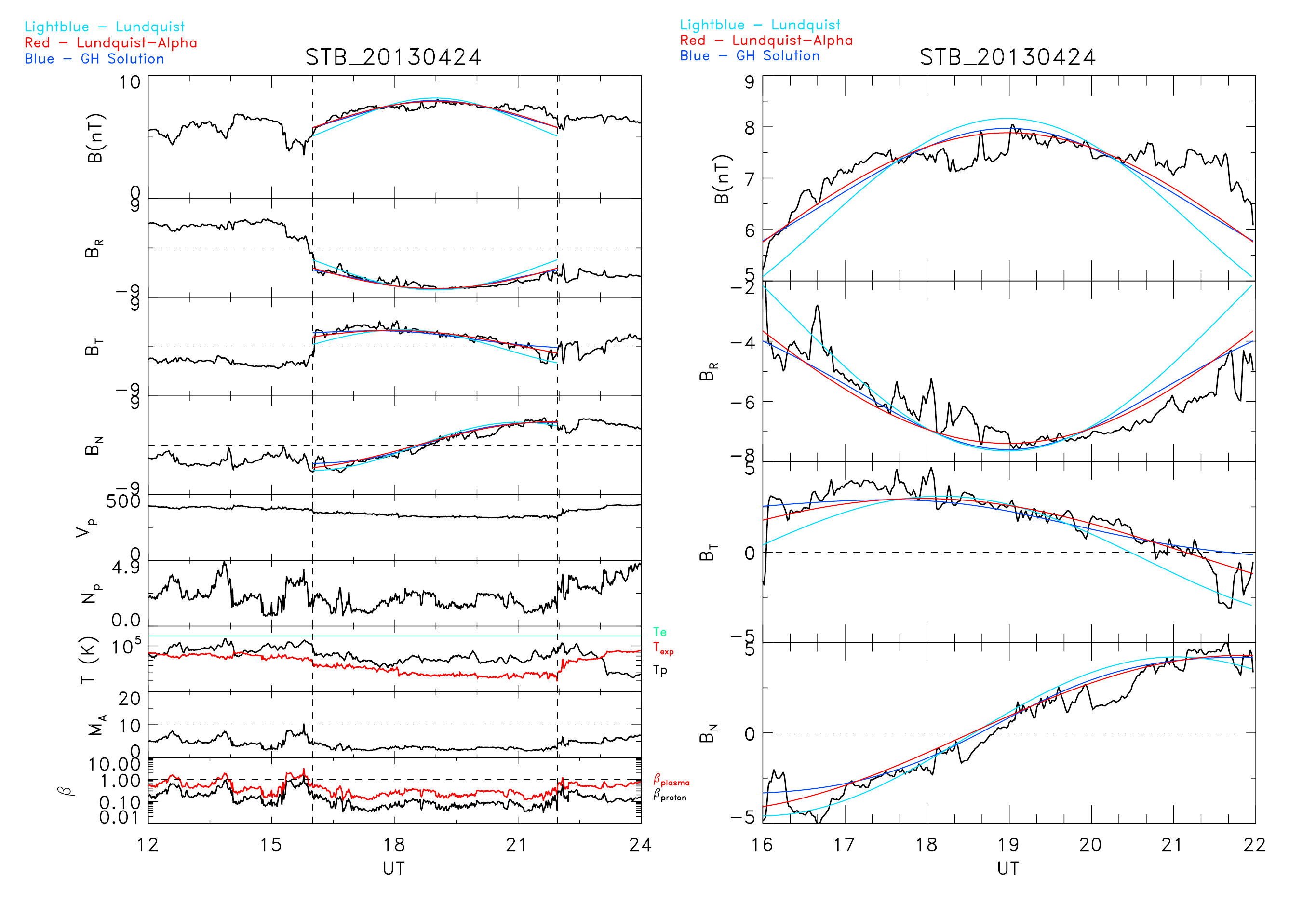}
\end{center} 
 \caption{Event 2: A SFR observed by STB on 24 April 2013. The SFR interval is bracketed by the two vertical lines. The format is similar to  that of Figure 1.}
\label{fig:event2 - analytical fitting results}
\end{figure}

\begin{figure}[!ht]
\begin{center} 
\includegraphics[width=4.8 in]{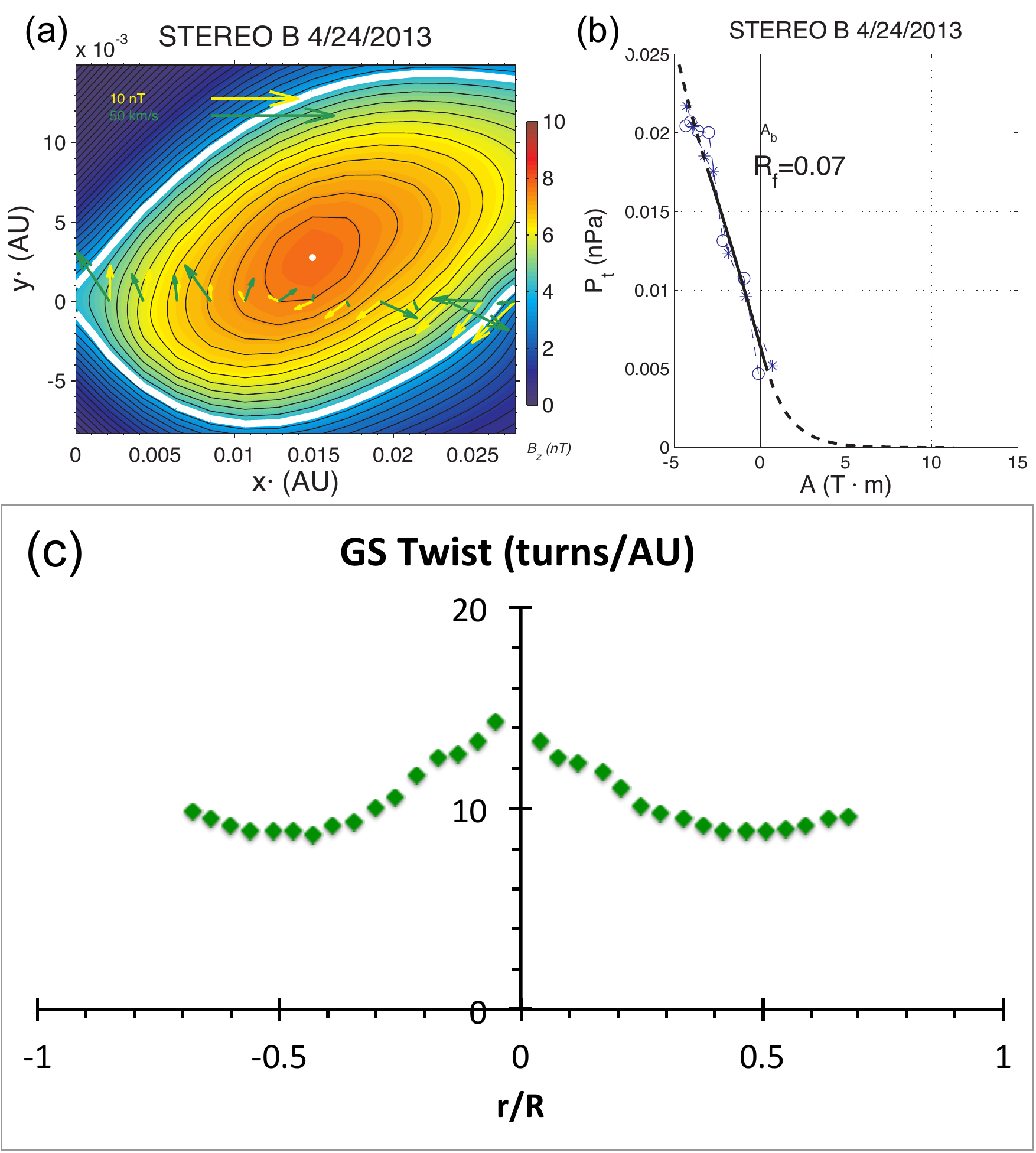}
\end{center} 
 \caption{Event 2: A SFR observed by STB on 24 April 2013. GS-reconstruction results: (a) magnetic field map; (b) fitting residue; (c) twist distribution from GS-reconstruction. Same format as Figure 2.}
\label{fig:event2 - GS results}
\end{figure}

The second example is from STB observations on 24 April 2013, the magnetic and plasma data for which are plotted in Figure 3. The SFR interval is from 16:00:00 to 21:58:00 UT, giving it a duration of 5.967 h, {\it i.e.}  longer than in the first example.  
This SFR has a total magnetic field strength which is higher than the surroundings, and smooth and large rotations in the magnetic field components, with $B_N$ changing polarity. In this event, the bulk speed $V_p$  decreases from the start (360 km/s) to the end (330 km/s) so there is some radial expansion. The expansion velocity is, however, only 15 km/s, which may be considered small, being just one-tenth of
average speed. $T_p$ is  higher than the expected temperature from solar wind expansion 
and much lower than $T_e$
, as often found in small FRs \citep{Yu2016}. $M_A$ is about 3, while both  $\beta_{proton}$ and $\beta_{plasma}$ are much less than 1.

From the fitting results which are plotted in the first four panels, both total magnetic field strength and its three components are fitted very well by all three models. The orientations obtained from the LA and GH models are very close. In addition, these two models also have the same magnetic field strength at the center. 
The impact parameter resulting from the  GH solution is the smallest of the three and not too far from 
that obtained from GS (0.39 {\it vs.} 0.28 $R$). However, the LA model  has the least normalized $\chi^2$ value (0.012 {\it vs.} 0.014 for GH).  Again, least-squares fitting results do not seem to suffice to select a preferred model.

Now we inquire into the twist {\it per} unit length. Figure 4 shows the GS-reconstruction results in the same format as Figure 2. 
The magnetic contour plot of this event shows a pronounced elliptical shape. 
The impact parameter, $p$, is smallest for the GH solution (= 0.393 $R$), but all models have $p < 0.5 R$.
The twist distribution for the in and out trajectories have practically the same values. The profile shows a marked decrease from the FR axis and then
appears to approach values of around 10 turns/AU.  This is  very close to the value of 10.2 turns/AU obtained from the GH solution. So this example suggests that GH is to be preferred.


\clearpage


\subsection{Event 3: Wind - 9 November 2013}

\begin{figure}[!ht]
\begin{center} 
\includegraphics[width=4.8 in]{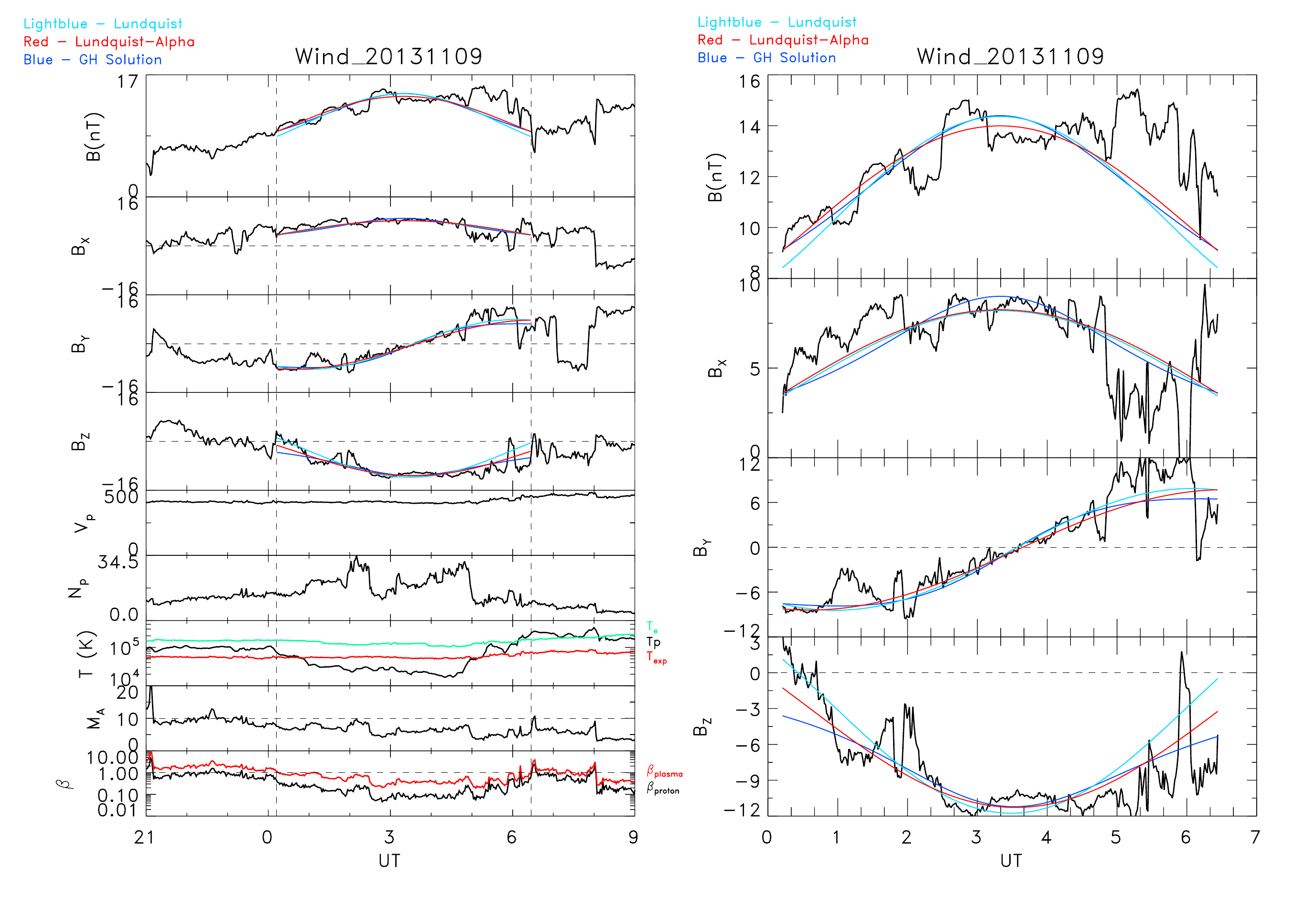}
\end{center} 
 \caption{Event 3: A SFR observed by {\it Wind} on 9 November 2013.  The SFR is bracketed by the two vertical lines.  Same format as in Figure 1 is used.}
\label{fig:event3 - analytical fitting results}
\end{figure}

\begin{figure}[!ht]
\begin{center} 
\includegraphics[width=4.8 in]{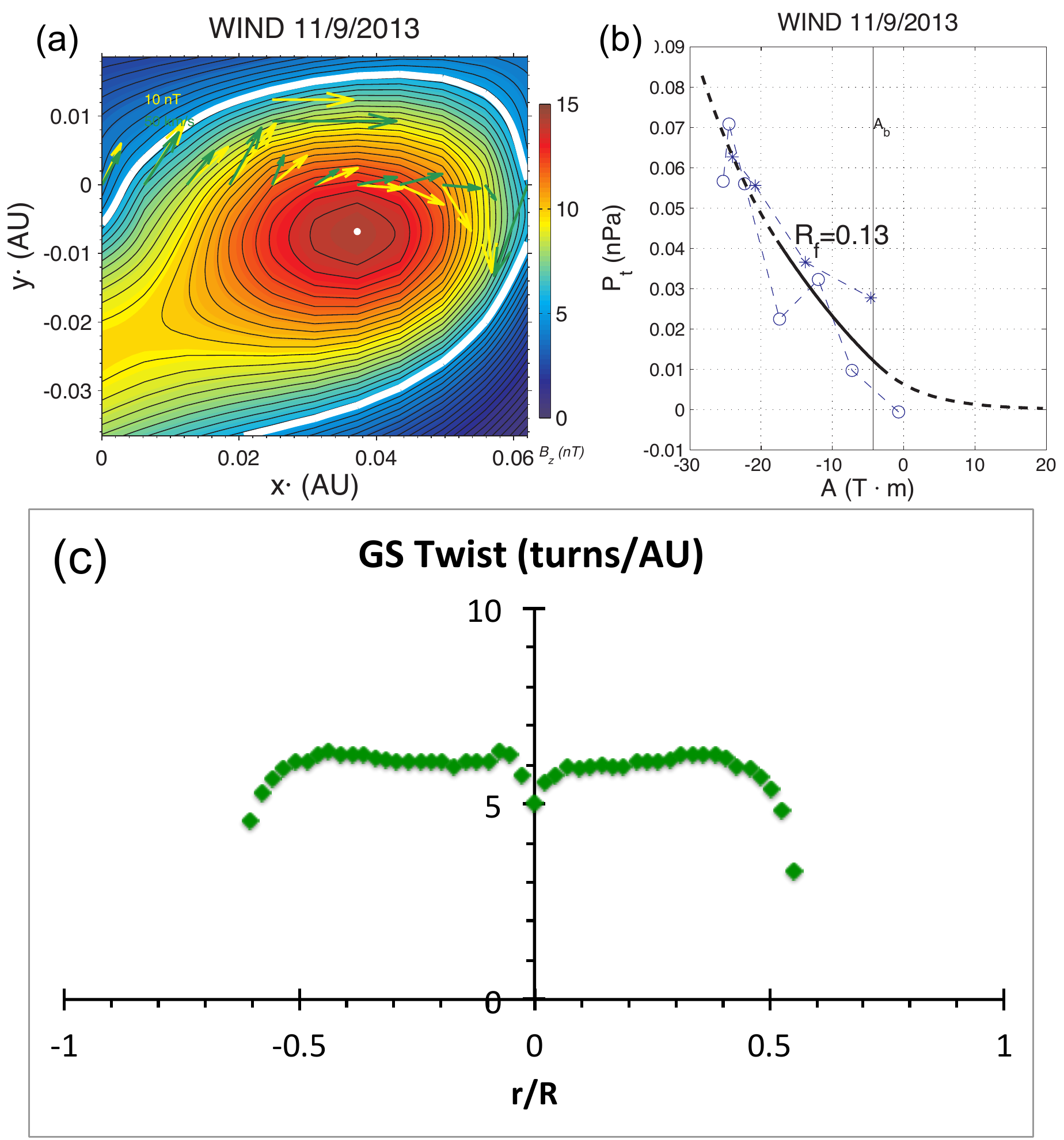}
\end{center} 
 \caption{Event 3: A SFR observed by {\it Wind} on 9 November 2013. GS-reconstruction results: (a) magnetic field map, (b) fitting residue, (c) twist distribution from GS-reconstruction. }
\label{fig:event3 - GS results}
\end{figure}

The third SFR event was observed by {\it Wind}  on 9 November 2013.  It lasted for 6.25 h (from 00:12:00 to 06:27:00 UT), making it the longest of the three cases. The panels in Figure 5 show the same quantities as Figures 1 and 3 but this time the magnetic field components are in the GSE coordinate system. This event has clear rotations in the field components and the magnetic 
field strength is enhanced. The  $V_p$ profile is very flat, indicative of a static structure. The proton temperature, $T_p$, is lower than the expected temperature 
and $T_e$. $M_A$, and $\beta_{proton}$, $\beta_{plasma}$ are small, even though the density is high (maximum $\sim$ 34.5 cm $^{-3}$) The fitted impact parameter ranges from 0.318 $R$ (GH) to  0.325 (LA) and 0.411 (L).

The analytical fits of the three models are very good for both total magnetic field strength as well as the field components (see the first four panels in Figure 5 and column nine in Table 3). The normalized $\chi^2$ of LA and GH solutions are close and  slightly smaller than the Lundquist solution.  Again from modeling alone it is hard to decide which FF model reproduces the field line geometry best.

GS-reconstruction results for this event are shown in Figure 6. It presents an elongated cross-section. The fitting residue of this event is 0.13.  The twist profile of this event shows that the input and output parts are generally consistent with each other. The twist profile is flat inside this flux rope, and decreases somewhat on the boundaries.  Such a profile is consistent with GH, which however overestimates the
result by a factor of 1.2 with respect to GS (7.1 {\it vs.} 5.8 turns/AU).  Note that for this longest of the three events $\tau$ is least.

In summary, analytical modeling is not enough to select between the models. There are some indications that the twist from GS is more
in line with that of the GH model.
Now we go to statistics to  obtain the general picture.

\section{Statistical Results}
\label{s:statistics}

We now consider the whole ensemble of 29 events and present statistical results.
Figure 7 shows the results on the twist profile for these events. Each event is shown by a different color. Figure 7a (top panel) shows the profile of the twist for the LA model as a function
of the radius ($r/R$) from closest approach (left) to the FR boundary. Clearly, for many of these FRs the twist varies considerably
and of course increases monotonically toward the FR boundary. In six of the events $\tau$ values exceed  60 turns/AU at the boundary. Figure 7b shows the distribution of twist values obtained from the GH model.
Twist values can be  fairly high, reaching up to 34, with $\tau$ in 10 events exceeding 20.  
66\% and in the range [5, 20]. 
Figure 7c (bottom panel) shows the twist distribution resulting from GS reconstruction.  We note that in two of the events the GS reconstruction failed. In those events whose twist at closest approach (left) exceeds $\sim$30, the trend is for the twist to
first decrease and then to reach steady values as the boundary of the reconstruction is approached (the thick white curve in Figures 2, 4, 6). In two other events, the twist did increase from the axis.
 For the others, the twists are relatively constant throughout.
This is clearly opposite to  the trend in the LA model (Figure 7a) and more in line with GH model (Figure 7b).  From here onwards we
thus restrict ourselves to discussing GH and GS results only.

We now give some average results for key parameters on the whole ensemble obtained from GH analysis. 
The top panel of Figure 8 shows the 
distribution of GH-fitted magnetic field strength on the FR axis.   Values range from 6 to 35 nT with
the highest frequency lying in the range 10--15 nT, which is significantly higher that typical solar wind values at 1 AU and fairly
comparable to typical values in MCs.
The middle panel gives the derived radii of these small FRs.  The distribution peaks in the  [0.005, 0.020 AU] range, which is of order 2--10$\%$ of typical MC radii (Lepping {\it et al.}, 1990, 2006).  
From Figure 9 of \inlinecite{Wang2016}, the radii of MCs discussed there are in the range [0.015, 0.17] AU. 
While there is thus a small overlap in sizes in the two groups, we are indeed 
extending the analysis to lower spatial scales than those 
considered so far. 
The bottom panel gives the twist distribution of these FRs. Twists range from 5--35 turns/AU. Thus these SFRs 
are magnetically dominated structures with highly twisted magnetic field lines. 

\begin{figure}[!ht]
\begin{center} 
\includegraphics[width=4.0 in]{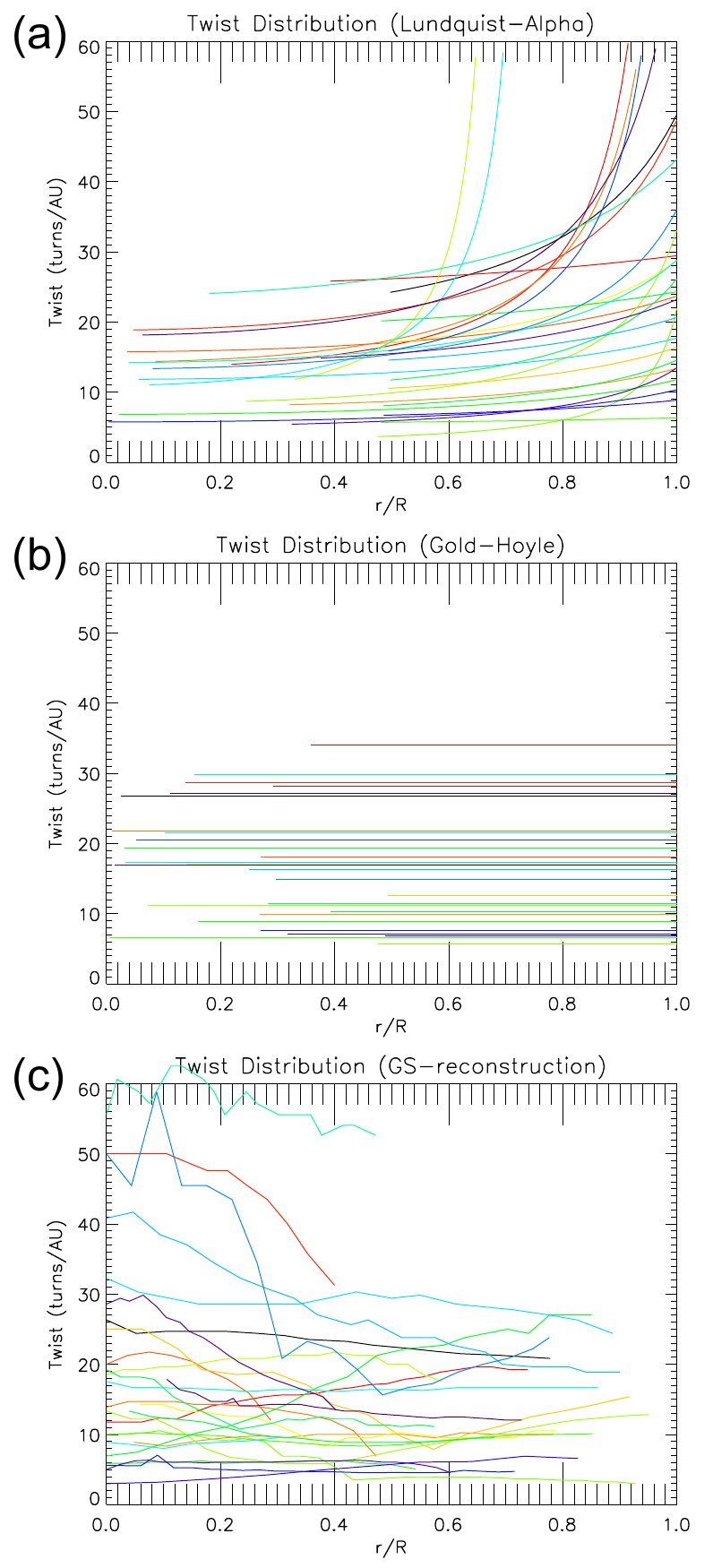}
\end{center} 
 \caption{Twist distribution of the set of 29 SFRs. (a) for LA, (b) for GH, (c) for GS. }
\label{fig: 29 twist vs. radius}
\end{figure}

\begin{figure}[!ht]
\begin{center} 
\includegraphics[width=4.4 in]{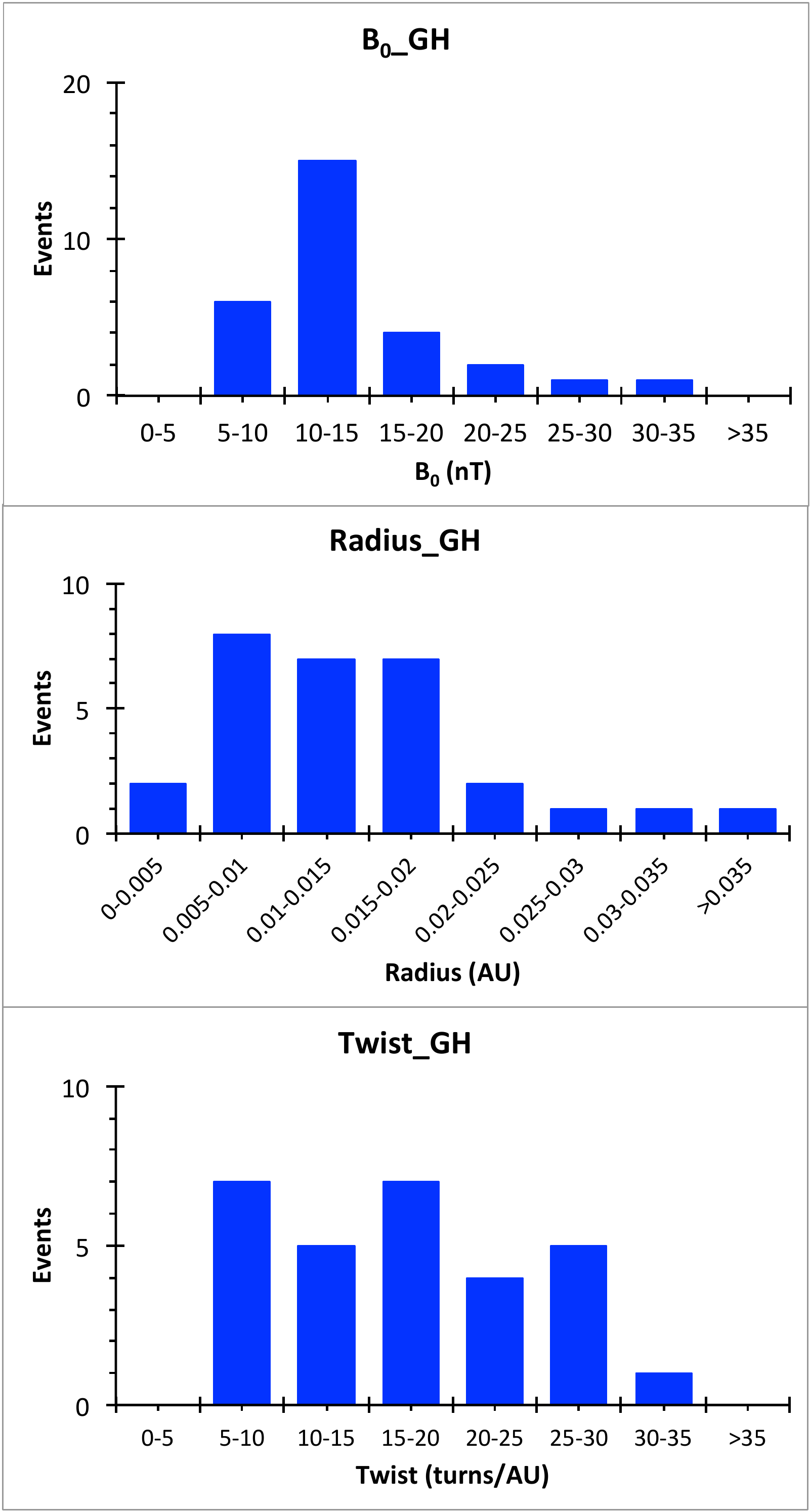}
\end{center} 
 \caption{Top panel: The distribution of the magnetic field strength on the axis.  Middle panel: distribution of the size (radius). Bottom panel: Distribution of twist in turns/AU. }
\label{fig: distributions of radius and twist}
\end{figure}

\begin{figure}[!ht]
\begin{center} 
\includegraphics[width=4.8 in]{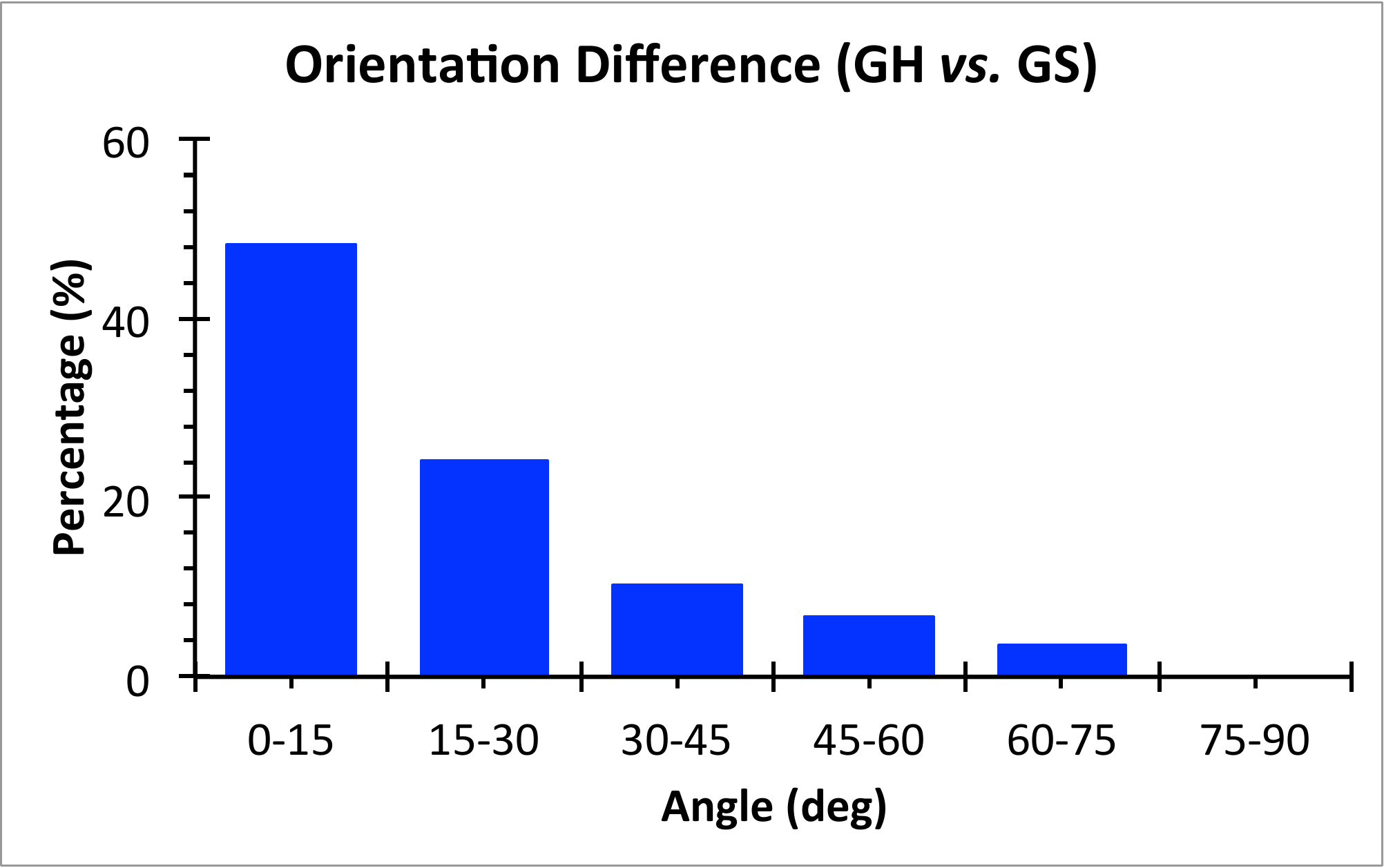}
\end{center} 
 \caption{The difference in orientation resulting from GH modeling and GS-reconstruction.}
\label{fig: orientations differences}
\end{figure}

\begin{figure}[!ht]
\begin{center} 
\includegraphics[width=4.8 in]{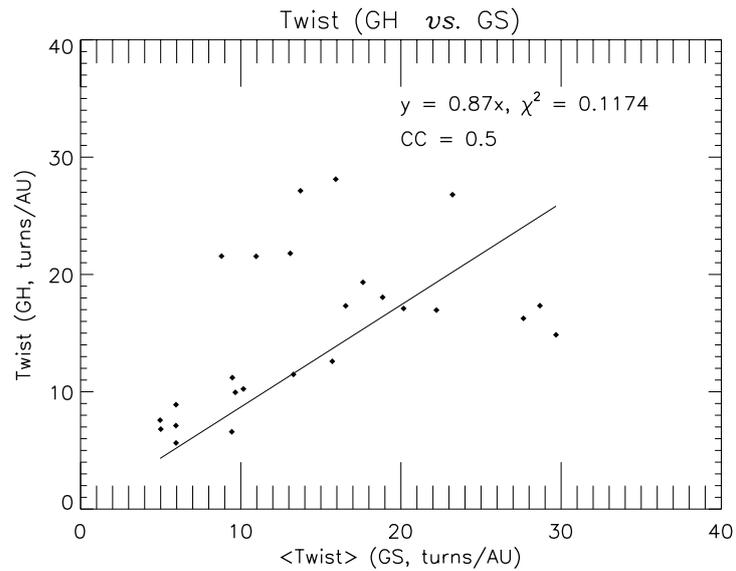}
\end{center} 
 \caption{Comparison of the twists obtained from the GH solution with the average twists from the GS-reconstruction.  }
\label{fig: twists differences}
\end{figure}

\begin{figure}[!ht]
\begin{center} 
\includegraphics[width=4.8 in]{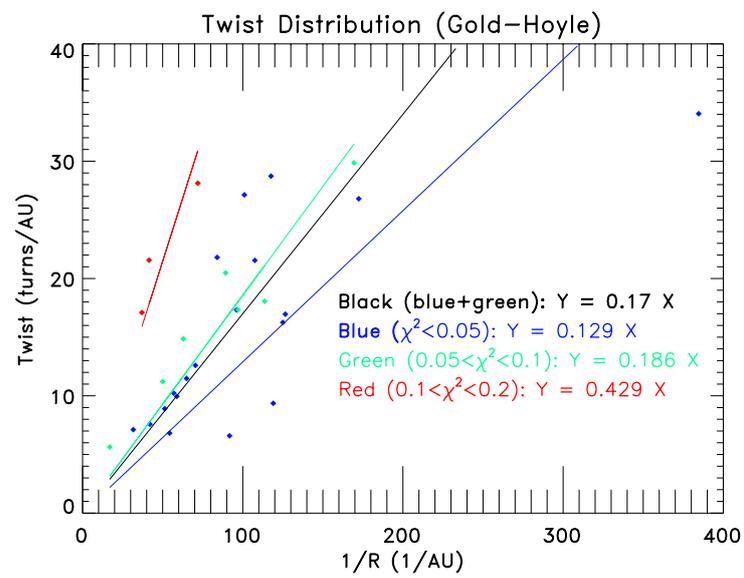}
\end{center} 
 \caption{Twists as a function of $1/R$ for the  29 SFRs. }
\label{fig:29 twist vs. 1/R}
\end{figure}

\begin{figure} [!ht]
\begin{center} 
\includegraphics[width=4.0 in]{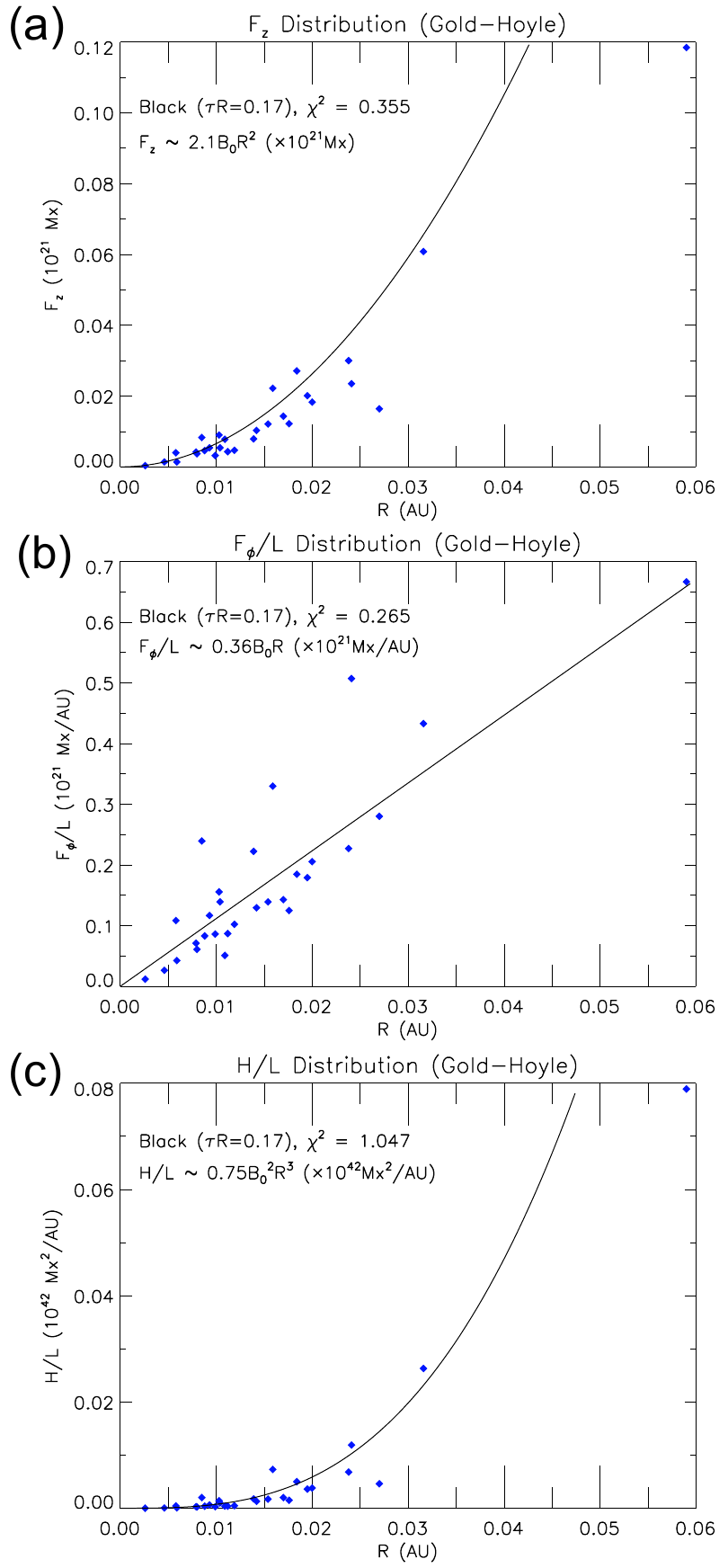}
\end{center} 
 \caption{Distributions of $F_z, F_{\phi}/L$, and $H/L$ quantities obtained from the GH solution. }
\label{fig: Fz, Fphi, H distributions}
\end{figure}

In Figures 9 and 10 we compare some results obtained from the GH and GS methods. Figure 9 shows the differences in the axis orientations. For
48$\%$ of the cases this difference is less than $15^{\circ}$ and for 72$\%$ it is less than $30^{\circ}$. This is a good
index of robustness.
A comparison of the twists obtained from GH with those from  GS is given in Figure 10 in the form of a scatter plot of the GH-derived twists {\it vs.} the averages of the twist inferred from GS reconstruction.  
The correlation coefficient of these twists is about 0.5. 
We use the IDL routine \textit{curvefit} and apply a straight line fit on the data, obtaining $\tau_{GH}  = 0.87\times \tau_{GS}$. 
According to this, the GH twist  for SFRs  underestimates the twist obtained from GS by a factor of $\sim$1.15. However, there
is obviously plenty of scatter, as reflected also in the values for the average and standard deviation of the ratio $\tau_{GH}/\tau_{GS}$ for which we obtain 1.37 $\pm$ 0.92. Thus, while a  correction cannot be ruled out, the
data is not so clear as to what it is.  In the next figures we shall thus ignore any correction.

\begin{figure}[!ht]
\begin{center} 
\includegraphics[width=4.8 in]{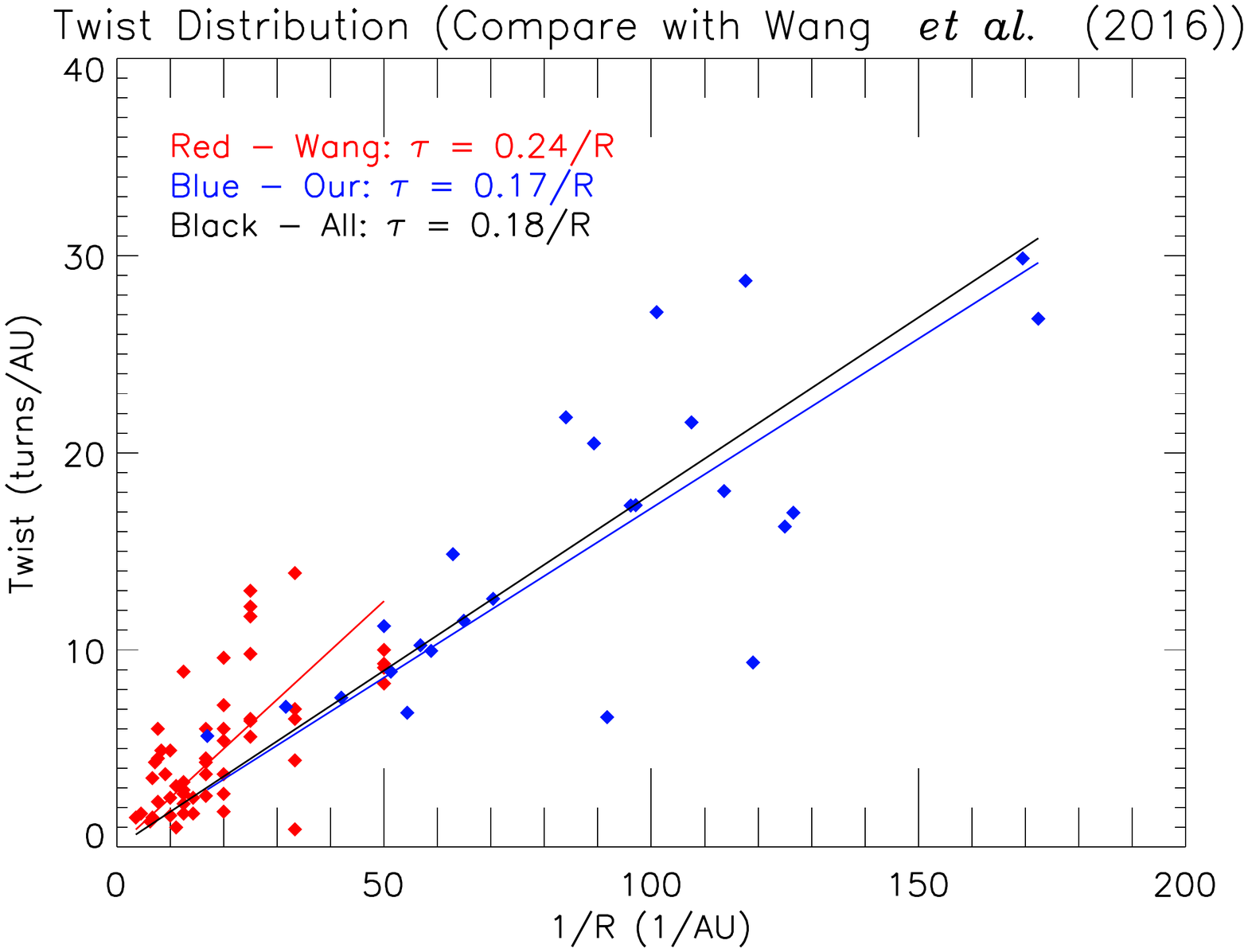}
\end{center} 
 \caption{The twist values for our SFRs (blue) and Wang {\it et al.} (2016) MC events as a function of $1/R$. For both sets, the best quality data are plotted. In both cases we show the twist as derived from GH-fitting. The black line is a fit to the two sets of data. }
\label{fig: Wang-US}
\end{figure}

We now come to the central aspect of this work, namely,  the dependence of $\tau$ on the radius of the tube, $R$. 
In Figure 11  we show the twist values as functions of $1/R$, as derived from GH fitting.  We group the data by
the normalized goodness-of-fit parameter $\chi^2$: blue for $\chi^2$ less than 0.05, green for $\chi^2$ between 0.05 and 0.1, and 
red for $\chi^2$ between 0.1 and 0.2. We then do a straight line fit using the IDL algorithm \textit{curvefit}. The gradients increase from blue to green to red. The black line gives a fit to the blue and green data points. The twist is inversely proportional to
$R$ as $\tau \sim 0.17/R$. 
Actually, we first did a power law fit ($\tau \sim R^{-\gamma}$) and obtained an index of 1.08. So assuming a linear fit is reasonable.

We now use this result to compute the axial (toroidal) and azimuthal (poloidal) magnetic fluxes, and the relative helicity {\it per} unit length. For the 
GH solution these are given by (for derivation see, {\it e.g.}, {\it Dasso et al.}, 2006).

\begin{eqnarray}
   \begin{cases}
    F_z = B_0 \pi {\textrm ln}(1 + 4\pi^2 \tau^2 R^2)/(4\pi^2\tau^2) , \text{axial flux}, 
    \\
    \frac{F_{\phi}}{L} = B_0 {\textrm ln}(1 + 4 \pi^2 \tau^2 R^2)/ (4 \pi \tau) , \text{azimuthal flux per unit length}, 
    \\
    \frac{H}{L} = \frac{\pi B_0^2}{16 \pi^3 \tau^3}[{\textrm ln}(1+4 \pi^2 \tau^2 R^2)]^2, \text{relative helicity per unit length}, 
  \end{cases}
\end{eqnarray}
where $\tau$ is in units of {\textrm turns/AU}.

Figure 12 shows the result. 
The black traces  are obtained using $\tau R$ = 0.17 as derived above. 
In this case, $F_z$ is a quadratic function of $R$ given by $F_z$ $\approx 2.1 B_0 R^2$ ($\times10^{21}$ {\textrm Mx}). 
The azimuthal flux {\it per} unit length is a linear function of $R$ varying approximately as $F_{\phi} \sim 0.36 B_0 R$ ($\times10^{21}$ {\textrm Mx/AU}). 
The relative helicity per unit length is a cubic in $R$, given by $H/L \approx 0.75 B_0^2 R^3$ ($\times10^{42}$ {\textrm Mx$^2$/AU}).
It is seen that the black traces fit the data well.

In summary, consideration of magnetic field line twist clearly shows that, even for small FRs, the Gold-Hoyle model represents the field line geometry of these SFRs better than the Lundquist solutions, both with fixed as well as with free but constant $\alpha$. Extrapolating our $\tau(R)$ relation to
the range of values of radii typically found in MCs (Lepping {\it et al.}, 2006), {\it i.e.} $\approx [0.06, 0.26]$ AU, we would get a twist range of $\tau \sim [0.7, 2.8]$. This is discussed further below.

\section{Discussion and Conclusions}
\label{s:discussion}

This study was motivated by previous efforts on MCs, detailed in the Section 1, where considerations of 
magnetic field line twist and length were employed to give further insight into the magnetic field structure of these
large flux rope transients. In particular, a model-independent method was used -- GS reconstruction -- to obtain 
the twist in turns/AU and compare it with that obtained from three analytical models. When this was done, one conclusion was that more agreement resulted with the GH than with the Lundquist solution ({\it e.g.,} Hu, Qiu, and Krucker, 2015; Wang {\it et al.}, 2016).   We extended this work into the small spatial scale regime, targeting interplanetary SFRs
during solar maximum conditions. This is the solar activity phase where force-free models are appropriate (Yu {\it et al.}, 2016) and this allows
a comparison with MC studies, which also rely on force-free conditions.
We recall that the radius range for our events was [0.006, 0.05] AU. This scale size overlapped slightly those covered in MCs studies. We  showed that even for SFRs the GH solution is a better model than the Lundquist solution, with or without a  prescribed, constant $\alpha$.  We now compare aspects obtained from our study of SFRs with those for MCs.

In comparison with MCs we obtained significantly higher twist values. Specifically, the twist values were in the range [5, 35] turns/AU, 
while those of MCs were [1, 14] \citep{Wang2016}.  
If we consider here only values obtained by least-squares fitting of the best quality data to the GH model, our twist distribution lies mainly in the range [5, 30] turns/AU (Figure 8) while for MCs \inlinecite{Wang2016}  obtained a twist distribution peaking in [1, 2] turns/AU.  Thus interplanetary SFRs
are dominated by a magnetic field which is strongly twisted. 
Whether some SFRs may be kink unstable is an issue to be considered in future work.
Both sets followed a linear relation of $\tau$ with $R^{-1}$, but the coefficients were different: 0.24 for MCs and 0.17 for SFRs.
There is thus some indication that, while the functional relations are the same, the numerical factor  may depend on the size of the transient. We examine this further.

In Figure 13 we show by red symbols the data from the Wang {\it et al.} (2016) study, obtained from GH fitting (see their Table 2, quality 1). There were 34 cases. In blue symbols we show our data, limiting ourselves to those  satisfying $\chi^2 \le 0.1$ (Figure 11). There are 24 cases.
 Note the small overlap around $1/R \approx$30-50. Note also that
the range of $1/R$ for SFRs is three times that of MCs.  The linear regression lines show the fitted functions. While the fits appear different, when we do a combined fit to both sets of data, shown by the black trace, we obtain $\tau = 0.18/R$.  
 Although there is some scatter, it is reasonably good, and the slope is almost 
the same as that given in our Figure 11 for SFRs. So we suggest that the relationship $\tau R$ = 0.18 is approximately valid for all interplanetary FRs, irrespective of size.

We can discuss some comparisons between some results obtained from the GH model with those from GS reconstruction.
We compared the twist in turns/AU.  We obtained a value of 1.37 $\pm$ 0.93 (mean and standard deviation). Since there is plenty of scatter around the average, we did not find it reasonable to include a correction to the GH-derived twist values.
Figure 14 shows in the top panel the axial fluxes (GH {\it vs.} GS) and in the bottom panel the azimuthal fluxes/AU.
Linear fits are given by the straight lines. For the axial fluxes, the GH values are lower than the GS values by a factor
of 0.9. For the azimuthal fluxes {\it per} AU, the GH values are 1.32 times those of derived from GS reconstruction.

\begin{figure}[!ht]
\begin{center} 
\includegraphics[width=4.8 in]{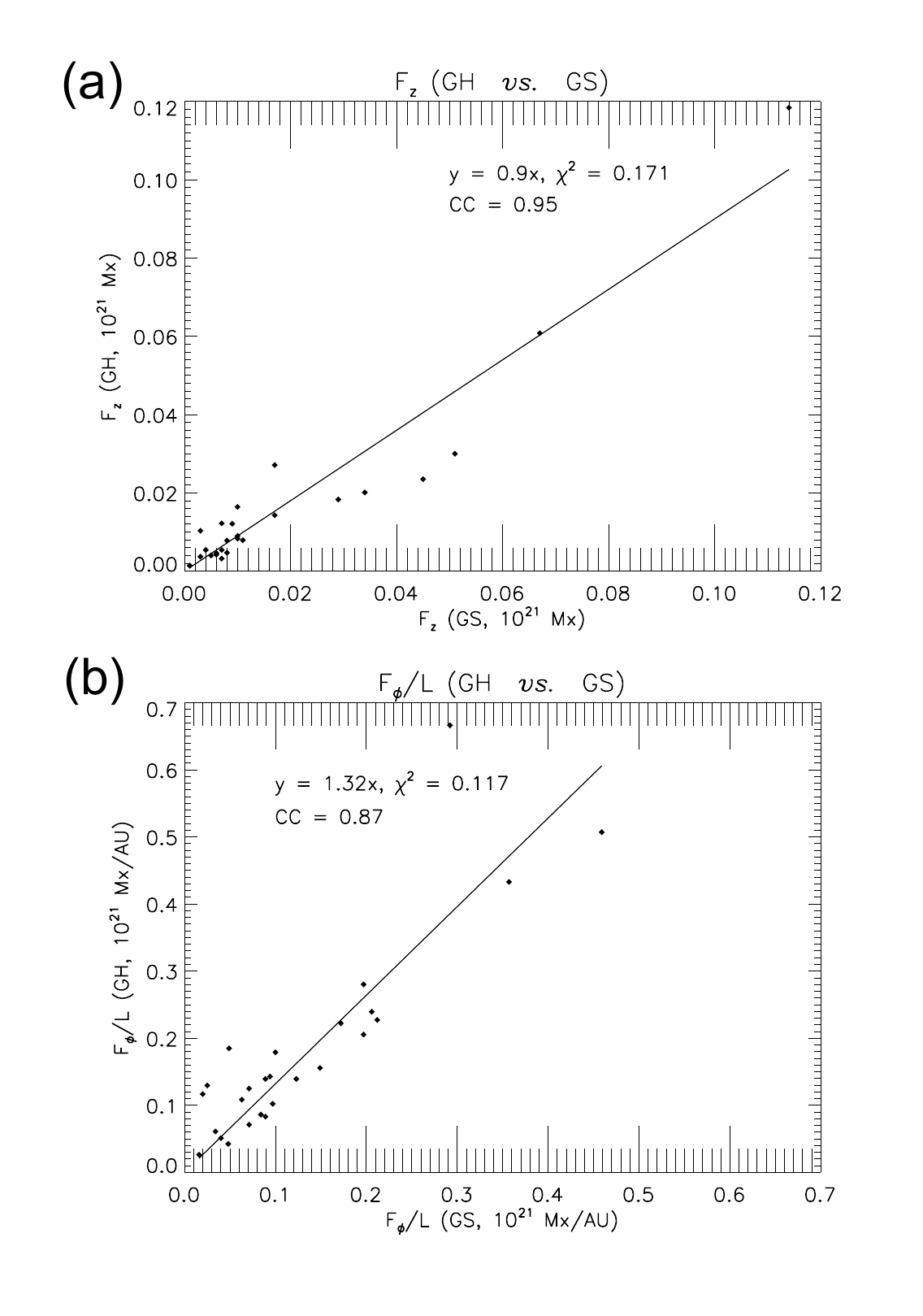}
\end{center} 
 \caption{The figure compares in the top panel the axial fluxes  and in the bottom panel the azimuthal fluxes/AU
derived from the GH model and GS reconstruction. The data points are fitted with a straight line. }
\label{fig: Fluxes}
\end{figure}

Wang {\it et al.} (2015) used a velocity-modified GH solution. This is warranted because the radial expansion velocity can be large for MCs. Thus, for  example, in his Figure 5, Demoulin (2010) showed a range of expansion velocities reaching up to 300 km s$^{-1}$, which is certainly a non-trivial 
fraction of the bulk velocity.  We assumed static structures and thus used the original GH model. How valid is this ? 
We find expansion in 15 out of the 29 cases.  A scatter plot of the radial expansion velocities is shown in Figure 15. All expansion velocities were below 25 km s$^{-1}$. This is only a small fraction of the  average speed, validating  our static assumption.

\begin{figure}[!ht]
\begin{center} 
\includegraphics[width=4.8 in]{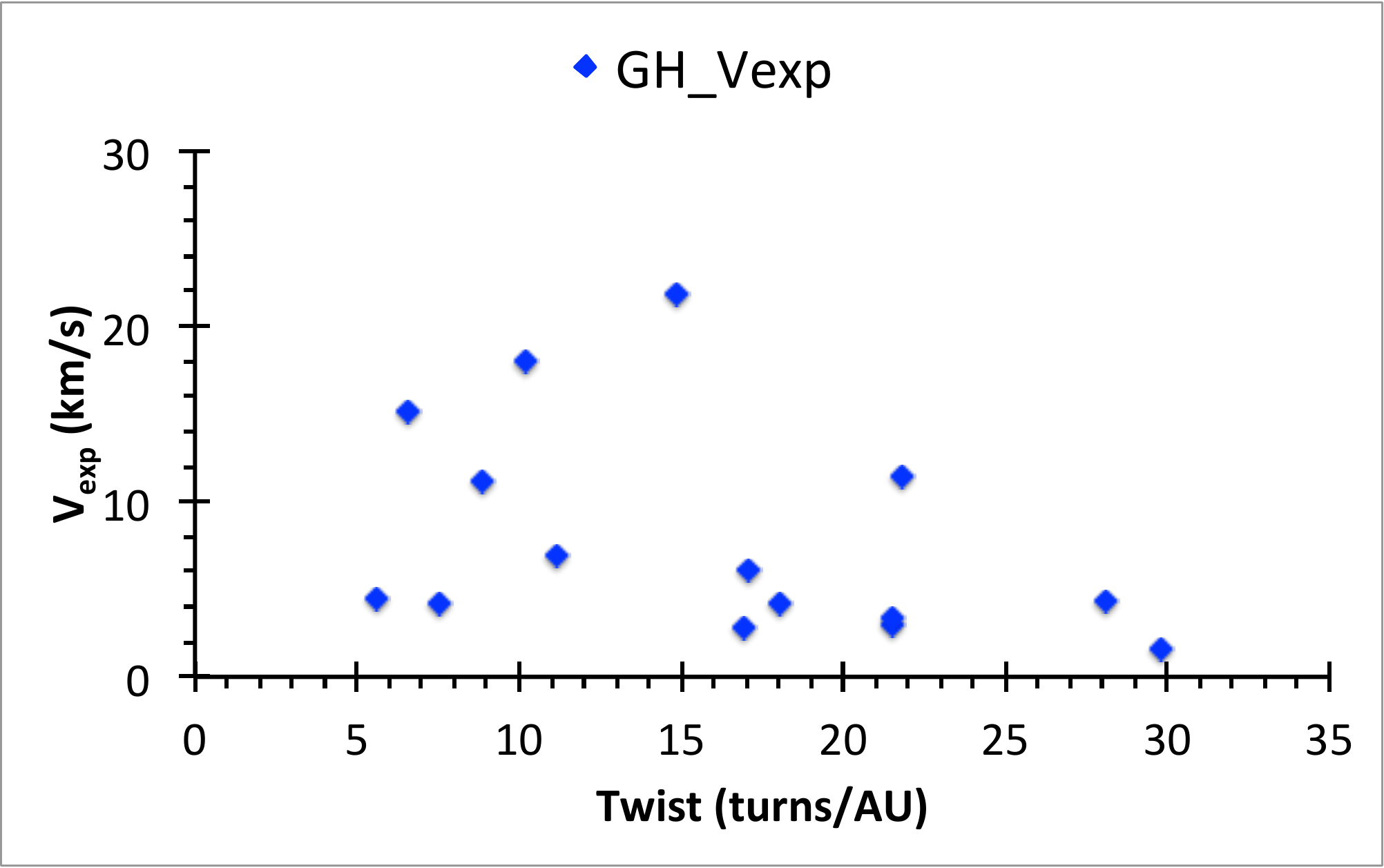}
\end{center} 
 \caption{Distribution of the expansion velocities of the SFRs (15/29). }
\label{fig: GH_Vexp}
\end{figure}

The magnetic field line twist was an essential ingredient in our study. Fitting quality alone is an ambiguous indicator of which model reflects the magnetic geometry of the transients best.
Thus in our case the  LA and GH had fairly similar values for the orientation of the flux rope and impact parameter, which admittedly for many considerations are the quantities that matter.  
Considerations of the twist distribution made a clean break: The LA solution was ruled out because it gave opposite trends to those obtained from GS.
Note that for MCs \inlinecite{Wang2016} showed that GH model is generally better than the Lundquist model, even from the least-squares fitting results.  A clear message is that the geometry of SFRs is not reproduced well by the linear force-free Lundquist model. If anything, when not constant the twist tends to peak near the axis of the FR not near its periphery.

The GH model agreement with the data is certainly not perfect.  In one case study, for example, the twist/unit length 
recovered from GS reconstruction indicated a flux rope which is more tightly wound near the axis. This is a trend which was often seen
when the twist was large. It suggests that for very high twist values no single model is sufficient to describe the field line structure of SFRs.

Using our linear relation between $\tau$ and $1/R$, we obtained pictures for the magnetic fluxes and relative helicity.  These quantities depend crucially on $\tau$. They are important because they also allow us to relate interplanetary flux ropes with a putative solar source (see {\it e.g} \inlinecite{Moestl2008}). Assume that $\tau R$ = 0.18 applies to both MCs and SFRs (see above). Assume also that
the most frequent $B_0$ for SFRs (10-15 nT) are also fairly representative of large MCs (Lepping {\it et al.}, 2006). The axial flux $F_z$, 
the azimuthal flux {\it per} unit length $F_\phi/L$ and the
relative helicity {\it per} unit length, $H/L$, depend on the radius as $R^2$, $R$ and $R^3$, respectively. Thus for example the axial fluxes in SFRs would
be  a factor of 100 lower than in MCs. In this respect we note that Mandrini {\it et al.} (2005) studied a small MC ($R$ = 0.016 AU, in our range) and linked it to a sigmoid eruption. They found that the axial flux was of order 100 less than an average MC. The tiny bipole on the Sun had also a factor of 100 less size than a classical active region.

In conclusion, 
in this article, we applied three analytical models  (Lundquist model, Lundquist-alpha model and GH model) and one numerical model (GS-reconstruction) on 29 SFRs. We found that the fitting quality of the GH solution is more or less the same as with Lundquist-alpha solution. 
Both models show better quality than the prevalent Lundquist solution (where $\alpha$ has a fixed value of $2.401/R$). 
The orientations and radii obtained  from the Lundquist-alpha and GH solutions are usually close, and more or less the same as when using the GS-reconstruction.   However, when we included the twist distribution, we found that the magnetic geometry of these
SFRs are better reproduced by the uniform-twist GH solution. The twist was much higher than typical of MCs. Thus we may think
of small solar wind flux ropes as structures dominated by a magnetic field which is strongly and uniformly twisted.
We also gave succinct expressions for the magnetic fluxes and the relative helicity which are important quantities when we search for
a possible solar origin of these events because they are conserved.

Finally, for SFRs, the force-free condition is  only valid for solar maximum conditions. A future work will address solar minimum conditions and non force-free models. This is very important since the frequency of SFRs increases at solar minimum and thus represents the  majority of flux ropes in the interplanetary medium (and associated geoeffects) when MCs are rare or absent.


\begin{table}   [!ht]
\caption{Analytical fitting results and GS-reconstruction of eight SFRs observed from STA}
\centering
\begin{footnotesize}
\begin{tabular}{c c c c c c c c c c c c c }
\hline
Events                & Model        & $\theta$ & $\phi$ & $B_0$ & H &  $R$  & $p/R$ & $\chi^2$ & $\tau$ & $F_z$ & $F_{phi}/L$ & $H/L$  \\
\hline
STA20120221    & L                & 34.9          & 252     & 30       & -     & 0.0071 & 0.5        & 0.023  & 27  & 0.0047   & 0.1979   & 0.0011       \\
02:38:00             & LA              & 36.1          & 262     & 27       & -    & 0.0073  & 0.5        & 0.016  & 30.2  & 0.0045  & 0.17     & 0.001     \\
04:00:00             & GH            & 32.8          & 298.9     & 24.3       & -     & 0.0058 & 0.025     & 0.021  & 26.8  & 0.004   & 0.1084  & 0.00044        \\
                           & GS            & 25             & 293     & 22.5      & -      & 0.0045 &  0           & 0.141  & 23.5   & 0.005  & 0.063    &             \\
\hline
STA20121228    & L                & -8          & 253.1     & 14.6      & +     & 0.0147  & 0.44        & 0.137  & 13  & 0.01   & 0.2009    & 0.0024       \\
05:02:00            & LA              & -6.8          & 273.1     & 13.4      & +     & 0.0142  & 0.219        & 0.141  & 46.4  & 0.0084   & 0.177    & 0.0018       \\
08:00:00            & GH            &  9.5          & 73.2     & 18      & +     & 0.0139  & 0.293        & 0.13  & 28.1  & 0.0079   & 0.2223    & 0.00176       \\
                          & GS            & -8           & 270.5     & 14.2      & +     & 0.013  & 0.2462        & 0.1166  & 16  & 0.011   & 0.172    &         \\
\hline
STA20130106    & L                & 5.7          & 96.5     & 9.3      & +     & 0.0033  & 0.314        & 0.084  & 58.2  & 0.0003   & 0.0284    & 1.08e-5       \\
18:40:00            & LA              & 6.3          & 106.5     & 8.5      & +     & 0.0033  & 0.392        & 0.02  & 27  & 0.003   & 0.015    & 8.88e-6       \\
19:28:00            & GH            & 5           & 129.9     & 8.5      & +     & 0.0026  & 0.358        & 0.02  & 34  & 0.00035   & 0.0119    & 4.14e-6       \\
                          & GS            &            &      &       &      &   &         &   &   &    &     &         \\
\hline
STA20130516   & L                & 58.6          & 288.9     &  30.8     & -     & 0.0085  & 0.173        & 0.042  & 22.6  & 0.0069   & 0.2427    & 0.002       \\
09:55:00            & LA             & 59.9          & 267.4     & 29.1      & -     & 0.0085  & 0.046        & 0.036  & 25  & 0.0065   & 0.214    & 0.0018       \\
11:30:00            & GH            & 59.2          & 285     & 32      & -     & 0.0085  & 0.139        & 0.03  & 28.7  & 0.0083   & 0.2394    & 0.002       \\
                          & GS            & 35          & 300     & 26.6      & -     & 0.0085  & 0.6588        & 0.1273  & 45  & 0.01   & 0.206    &        \\
\hline
STA20130726   & L                & 47          & 170.1     & 12.5      & -     & 0.0066  & 0.178        & 0.082  & 29.1  & 0.0017   & 0.0764    & 1.55e-4      \\
13:22:00            & LA             & 63.2          & 170.1     & 11.5      & -     & 0.0078  & 0.0366        & 0.06  & 17.9  & 0.0022   & 0.067    & 0.00022       \\
15:02:00            & GH            & 76.2          & 170.1     & 12.2      & -     & 0.0088  & 0.27        & 0.059  & 18.1  & 0.0046   & 0.0831    & 0.00038       \\
                          & GS            & 76.7          & 170     & 12.6      & -     & 0.008  & 0.4625        & 0.1711  & 18.6  & 0.006   & 0.089    &        \\
\hline
STA20140430   & L                & 20.8          & 303.9     & 9.4      & -     & 0.0134  & 0.113        & 0.03  & 14.3  & 0.0053   & 0.117    & 7.4e-4       \\
13:02:00           & LA              & 20.2          & 306.5     & 9.3      & -     & 0.013  & 0.086        & 0.029  & 26.3  & 0.0049   & 0.111    & 0.00066       \\
16:45:00           & GH            & 18.4          & 312.5     & 9.6      & -     & 0.0119  & 0.009        & 0.036  & 21.8  & 0.0047   & 0.1022    & 0.00048       \\
                         & GS            & 18          & 312     & 7.8      & -     & 0.015  & 0.02        & 0.0803  & 13.1  & 0.008   & 0.097    &        \\
\hline
STA20140506  & L                & 46.7          & 187.5     & 12.2      & -     & 0.0146  & 0.489        & 0.032  & 13.1  & 0.0082   & 0.1666    & 0.0016       \\
17:30:00           & LA             & 67.6          & 197     & 10.3      & -     & 0.017  & 0.321        & 0.02  & 9.5  & 0.0095   & 0.14    & 0.0019       \\
20:55:00           & GH            & -69          & 18.6     & 10.6      & -     & 0.017  & 0.269        & 0.019  & 10  & 0.0143   & 0.1428    & 0.002       \\
                         & GS            & 66          & 205     & 9.5      & -     & 0.0125  & 0.368        & 0.1654  & 9.8  & 0.017   & 0.094    &        \\
\hline
STA20140728  & L                & 55.5          & 163.9     & 11.3      & +     & 0.0137  & 0.437        & 0.074  & 13.9  & 0.0067   & 0.1444    & 0.0012       \\
03:38:00           & LA             & 55.5          & 166.5     & 10.6      & +     & 0.0142  & 0.5        & 0.039  & 12.1  & 0.0067   & 0.119    & 0.0011       \\
06:20:00           & GH            & -55.5          & 345.8     & 11.3      & +     & 0.0142  & 0.494        & 0.036  & 12.6  & 0.0103   & 0.1295    & 0.0013       \\
                         & GS            & 28          & 159     & 9.7      & +     & 0.0048  & 0.6875        & 0.2006  & 15.7  & 0.003   & 0.025    &        \\
\hline
\end{tabular}
\end{footnotesize}
\\
  \small Note: 
   Model - The force-free model we used (L - Lundquist solution with prescribed alpha, LA - Lundquist with non-fixed alpha, GH - Gold-Hoyle solution, GS - GS-reconstruction). $\theta (^{\circ})$ is the latitude angle. $\phi (^{\circ})$ is the longitude angle. $B_0$ ({\textrm nT}) is the magnetic field strength along the axis. H is the handedness of the magnetic field ("-" means left-handed, "+" means right-handed). $R$ ({\textrm AU}) is the radius of the flux rope cross-section. $p/R$ is the impact parameter. $\chi^2$ is the normalized chi-square from the fitting (for GS-reconstruction, fitting residue ($R_f$) is presented). $\tau$ ({\textrm turns/AU}): For L and LA, we obtained the average values of magnetic field twists along the axis. For GH, it is obtained from fitting. For GS it is the average twist value derived from the graphic approach. $F_z$ is axial magnetic flux. $F_{\phi}/L$ is azimuthal magnetic flux {\it per} unit length. $H/L$ is the relative helicity {\it per} unit length.
\label{tab: fitting results - STA}
\end{table}

%

\begin{table}   [!ht]
\caption{ Analytical fitting results and GS-reconstruction of ten SFRs observed from STB}
\centering
\begin{footnotesize}
\begin{tabular}{c c c c c c c c c c c c c }
\hline
Events                & Model        & $\theta$ & $\phi$ & $B_0$ & H &  $R$  & $p/R$ & $\chi^2$ & $\tau$ & $F_z$ & $F_{phi}/L$ & $H/L$  \\
\hline
STB20120220    & L                & -0.17          & 89.9     & 14.4      & -     & 0.0091  & 0.44        & 0.065  & 21.1  & 0.0037   & 0.122    & 0.0005       \\
21:17:00            & LA              & -2.1          & 89.9     & 13.5      & -     & 0.0093  & 0.483        & 0.037  & 19.9  & 0.0037   & 0.103    & 0.00052       \\
23:30:00            & GH            & -2.1           & 89.9     & 14.8      & -     & 0.0093  & 0.483        & 0.032  & 21.5  & 0.0054   & 0.1167    & 0.00063       \\
                          & GS            & 2           & 126.5     & 12      & -     & 0.0052  & 0.4231        & 0.0933  & 11.1  & 0.004   & 0.02    &         \\
\hline
STB20120419    & L                & 70.3          & 113.6     & 9.5      & +     & 0.029  & 0.354        & 0.127  & 6.6  & 0.0252   & 0.2577    & 0.0078       \\
05:22:00            & LA              & 70.3          & 53.6     & 12.6      & +     & 0.0285  & 0.33        & 0.072  & 12.3  & 0.0319   & 0.317    & 0.0127       \\
09:55:00            & GH            & 70.3           & 113.6     & 11.8      & +     & 0.027  & 0.141        & 0.129  & 17.1  & 0.0164   & 0.2801    & 0.0046       \\
                          & GS            & 33.2           & 157.8     & 11      & +     & 0.018  & 0.1722        & 0.0704  & 20.4  & 0.01   & 0.197    &         \\
\hline
STB20120507    & L                & -54.7          & 344.3     & 12.5      & +     & 0.0179  & 0.32        & 0.08  & 10.7  & 0.0124   & 0.207    & 0.0031       \\
12:22:00            & LA              & -74.2          & 225.2     & 11.9      & +     & 0.0206  & 0.245        & 0.075  & 12.5  & 0.0158   & 0.22    & 0.0043       \\
17:12:00            & GH            & -73           & 319.5     & 11.9      & +     & 0.02  & 0.072        &0.074   & 11.2  & 0.0183   & 0.2054    & 0.0038       \\
                          & GS            & -70           & 310     & 10.9      & +     & 0.0185  & 0.3027        & 0.1995  & 9.5  & 0.029   & 0.197    &         \\
\hline
STB20120520    & L                & 2.6          & 26.1     & 8.96      & +     & 0.026  & 0.464        & 0.138  & 7.4  & 0.0188   & 0.2156    & 0.0049       \\
0520/19:25:00    & LA              & -8.4          & 266.1     & 9.7      & -     & 0.059  & 0.476        & 0.098  & 5.9  & 0.1053   & 0.523    & 0.0673       \\
0521/05:22:00    & GH            & 8.4           & 86.1     & 12.6      & -     & 0.059  &0.476         & 0.083  & 5.6  & 0.1184   & 0.6662    & 0.0789       \\
                          & GS            & -5.5           & 312.8     & 8.4      & -     & 0.043  & 0.2628        & 0.1948  & 6.8  & 0.114   & 0.292    &         \\
\hline
STB20120614    & L                & 2.5          & 273.8     & 12.1      & +     & 0.0149  & 0.447        & 0.068  & 12.9  & 0.0084   & 0.1674    & 0.0017       \\
05:48:00            & LA              & 2.3          & 290.8     & 10.5      & +     & 0.0142  & 0.481        & 0.006  & 5.9  & 0.0066   & 0.0772    & 0.0011       \\
08:27:00            & GH            & 2           & 305.1     & 10.2      & +     & 0.0109  & 0.006        & 0.006  & 6.6  & 0.0078   & 0.051    & 0.0004       \\
                          & GS            & -1           & 300     & 10.5      & +     & 0.006  & 6.67e-2        & 0.1885  & 9.4  & 0.008   & 0.04    &         \\
\hline
STB20130124    & L                & 12.4          & 140.6     & 13.2      & +     & 0.0149  & 0.468        & 0.046  & 12.8  & 0.0093   & 0.1842    & 0.0021       \\
0124/21:25:00    & LA              & 18.9          & 98.7     & 11      & +     & 0.0199  & 0.021        & 0.036  & 8.1  & 0.0137   & 0.174    & 0.0034       \\
0125/01:38:00    & GH            & 18.3           & 107.9     & 11.4      & +     & 0.0195  & 0.16        & 0.036  & 8.9  & 0.0201   & 0.179    & 0.0036       \\
                          & GS            & 15           & 110     & 9.8      & +     & 0.0155  & 0.4581        & 0.1418  & 6  & 0.034   & 0.1    &         \\
\hline
STB20130223    & L                & 51.6          & 176.2     & 12.9      & +     & 0.0042  & 0.48        & 0.051  & 46  & 0.0007   & 0.0501    & 4.2e-5       \\
12:02:00            & LA              & 77.5          & 169.3     & 11      & +     & 0.0052  & 0.482        & 0.011  & 21.4  & 0.00093   & 0.0362    & 5.92e-5       \\
13:02:00            & GH            & -82.3           & 210     & 10.5      & +     & 0.0046  & 0.0042        & 0.012  & 19.3  & 0.0014   & 0.0264    & 3.6e-5       \\
                          & GS            & 82.8           & 81.4     & 10.4      & +     & 0.0027  & 0.1852        & 0.0567  & 17.8  & 0.001   & 0.016    &         \\
\hline
STB20130424    & L                & 16.4          & 126.2     & 9.5      & -     & 0.022  & 0.445        & 0.024  & 8.5  & 0.015   & 0.1983    & 0.0036       \\
16:00:00            & LA              & 14.1          & 134.8     & 8.7      & -     & 0.0204  & 0.486        & 0.0115  & 9.6  & 0.0113   & 0.148    & 0.0023       \\
21:58:00            & GH            & -12.9           & 320.2     & 8.7      & -     & 0.0176  & 0.393        & 0.0136  & 10.2  & 0.0122   & 0.1249    & 0.0015       \\
                          & GS            & -2.7           & 147.4     & 7.8      & -     & 0.01  & 0.28        & 0.0713  & 10.2  & 0.007   & 0.071    &         \\
\hline
STB20140118    & L                & 38.8          & 133.2     & 12.6      & -     & 0.014  & 0.428        & 0.022  & 13.7  & 0.0078   & 0.1642    & 0.0015       \\
06:12:00            & LA              & 43          & 126     & 12      & -     & 0.0156  & 0.498        & 0.013  & 14.9  & 0.0091   & 0.162    & 0.0019       \\
09:50:00            & GH            & 47.6           & 104.8     & 11.2      & -     & 0.0154  & 0.283        & 0.017  & 11.5  & 0.0121   & 0.1392    & 0.0017       \\
                          & GS            & 38.7           & 120     & 11.1      & -     & 0.012  & 0.275        & 0.132  & 12.8  & 0.009   & 0.123    &         \\
\hline
STB20140523    & L                & -48.7          & 257.5     & 9.2      & -     & 0.0059  & 0.126        & 0.073  & 32.7  & 0.001   & 0.0502    & 6e-5       \\
13:00:00            & LA              & -46.7          & 257.5     & 8.8      & -     & 0.0059  & 0.18        & 0.052  & 28.8  & 0.00096   & 0.042    & 5.54e-5       \\
14:35:00            & GH            & -46.9           & 257.5     & 9      & -     & 0.0059  & 0.155        & 0.052  & 29.9  & 0.0014   & 0.0425    & 6.06e-5       \\
                          & GS            & -39.8           & 327.3     & 10      & -     & 0.0053  & 0.6415        & 0.0881  & 57.4  & 0.001   & 0.048    &         \\
\hline
\end{tabular}
\end{footnotesize}
\\
  \small Note: Same parameters as in Table 1 but for STB. 
\label{tab: fitting results - STB}
\end{table}

%

\begin{table}   [!ht]
\caption{{\bf Analytical fitting results and GS-reconstruction of eleven SFRs observed from {\it Wind}}}
\centering
\begin{footnotesize}
\begin{tabular}{c c c c c c c c c c c c c }
\hline
Events                & Model        & $\theta$ & $\phi$ & $B_0$ & H &  $R$  & $p/R$ & $\chi^2$ & $\tau$ & $F_z$ & $F_{phi}/L$ & $H/L$  \\
\hline
Wind20000916   & L                & -72.8          & 308.2     & 17.1      & +     & 0.0104  & 0.18        & 0.054  & 18.3  & 0.0058   & 0.1662    & 0.0012       \\
20:58:00            & LA              & -74.8          & 248.2     & 16      & +     & 0.0104  & 0.039        & 0.042  & 17.5  & 0.0054   & 0.139    & 0.001       \\
23:10:00            & GH            & -75.1           & 248.2     & 16.3      & +     & 0.0104  & 0.033        & 0.045  &17.3   & 0.0054   & 0.1392    & 0.001       \\
                          & GS            & -63.3           & 185     & 16.4      & +     & 0.0065  & 0.4615        & 0.2  & 16.6  & 0.007   & 0.089    &         \\
\hline
Wind20001104    & L                & -67.6          & 190.6     & 17.6      & -     & 0.0224  & 0.043        & 0.152  & 8.6  & 0.0276   & 0.3663    & 0.0122       \\
02:30:00            & LA              & -73.7          & 200     & 20.7      & -     & 0.0233  & 0.075        & 0.129  & 26.9  & 0.035   & 0.445    & 0.019       \\
07:30:00            & GH            & -83.3           & 202.9     & 25      & -     & 0.0241  &  0.102       & 0.137  & 21.6  & 0.0235   & 0.507    & 0.0119       \\
                          & GS            & -48           & 240     & 21      & -     & 0.02  & 0        & 0.1792  & 8.9  & 0.045   & 0.459    &         \\
\hline
Wind20010322    & L                & 5.7          & 323.9     & 21.6      & -     & 0.0072  & 0.486        & 0.108  & 26.6  & 0.0035   & 0.1447    & 0.0006       \\
14:40:00            & LA              & 11.8          & 281.4     & 17.5      & -     & 0.0104  & 0.0565        & 0.092  & 13.4  & 0.0059   & 0.135    & 0.0012       \\
17:10:00            & GH            & -10.3           & 105.4     & 18.6      & -     & 0.0103  & 0.105        & 0.09  & 17.3  & 0.009   & 0.1556    & 0.0014       \\
                          & GS            & 20           & 290     & 20.8      & -     & 0.008  & 0        & 0.1375  & 28.7  & 0.01   & 0.149    &         \\
\hline
Wind20010826    & L                & -0.86          & 260     & 12.8      & -     & 0.009  & 0.445        & 0.05  & 21.3  & 0.0032   & 0.1072    & 0.0004       \\
07:35:00            & LA              & -0.7          & 275.1     & 11.4      & -     & 0.0094  & 0.494        & 0.025  & 16.3  & 0.0031   & 0.0808    & 0.00038       \\
09:10:00            & GH            & -0.8           & 288.5     & 10.9      & -     & 0.008  & 0.25        & 0.026  & 16.3  & 0.0037   & 0.0609    & 0.00023       \\
                          & GS            & 12           & 298     & 11.3      & -     & 0.0034  & 0.0588        & 0.1279  & 27.6  & 0.003   & 0.034    &         \\
\hline
Wind20031013    & L                & 37.1          & 105.5     & 24.6      & -     & 0.018  & 0.439        & 0.082  & 10.5  & 0.0259   & 0.419    &  0.013      \\
14:50:00            & LA              & 26.9          & 141.4     & 21.3      & -     & 0.0121  & 0.081        & 0.078  & 17.7  & 0.0098   & 0.226    & 0.0028       \\
18:58:00            & GH            & -34.6           & 302.2     & 23.7      & -     & 0.0159  & 0.297        & 0.082  & 14.9  & 0.0222   & 0.3298    & 0.0073       \\
                          & GS            & 76.6           & 284.5     & 23.4      & -     & 0.0015  & 0.1333        & 0.0818  & 28.5  &    &     &         \\
\hline
Wind20120407    & L                & -7          & 72.3     & 10.1      & -     & 0.0141  & 0.496        & 0.058  & 13.6  & 0.0063   & 0.1327    & 0.001       \\
09:18:00            & LA              & -7.4          & 74.7     & 9.9      & -     & 0.0143  & 0.495        & 0.058  & 29.4  & 0.0064   & 0.131    & 0.001       \\
12:30:00            & GH            & 7.3           & 300.3     & 9      & -     & 0.0112  & 0.052        & 0.055  & 20.5  & 0.0043   & 0.087    & 0.00037       \\
                          & GS            &            &      &       &      &   &         &   &   &    &     &         \\
\hline
Wind20120802    & L                & -1.5          & 237.7     & 13.1      & -     & 0.023  & 0.439        & 0.046  & 8.4  & 0.0215   & 0.2791    & 0.0072       \\
14:28:00            & LA              & -0.7          & 272.9     & 11.1      & -     & 0.024  & 0.002        & 0.034  & 6.8  & 0.0205   & 0.218    & 0.0062       \\
19:00:00            & GH            & -0.3           & 289.8     & 11.7      & -     & 0.0238  & 0.27        & 0.036  & 7.6  & 0.03   & 0.2271    & 0.0068       \\
                          & GS            & -1           & 285     & 9.9      & -     & 0.038  & 0.4474        & 0.151  & 5  & 0.051   & 0.212    &         \\
\hline
Wind20130620    & L                &  -19.7         & 268.2     & 17      & +     & 0.019  & 0.465        & 0.055  & 9.9  & 0.0199   & 0.306    & 0.0073       \\
12:38:00            & LA              & -19.7          & 290.4     & 14.7      & +     & 0.0185  & 0.485        & 0.035  & 7.3  & 0.0157   & 0.193    & 0.0048       \\
16:28:00            & GH            & -19.7           & 291.6     & 14.6      & +     & 0.0184  & 0.489        & 0.038  & 6.8  & 0.0271   & 0.1847    & 0.005       \\
                          & GS            & -2.83           & 293     & 13.6      & +     & 0.0075  & 0.4133        & 0.1047  & 4.8  & 0.017   & 0.049    &         \\
\hline
Wind20131109    & L                & -81.9          & 311.3     & 16.2      & -     & 0.034  & 0.411        & 0.04  & 5.7  & 0.0579   & 0.5096    & 0.0355       \\
00:12:00            & LA              & -71.6          & 338.3     & 14.8      & -     & 0.0312  & 0.325        & 0.035  & 7.1  & 0.0451   & 0.401    & 0.0233       \\
06:27:00            & GH            & 73.9           & 154.7     & 15.8      & -     & 0.0316  & 0.318        & 0.037  & 7.1  & 0.0608   & 0.4328    & 0.0263       \\
                          & GS            & -75           & 349     & 14.4      & -     & 0.0257  & 0.2646        & 0.1267  & 5.8   & 0.067   & 0.357    &         \\
\hline
Wind20131214    & L                & -17          & 252.9     & 15.5      & +     & 0.0095  & 0.474        & 0.05  & 20.2  & 0.0044   & 0.1369    & 0.0007       \\
02:58:00            & LA              & -17.6          & 276.1     & 13.3      & +     & 0.0094  & 0.376        & 0.039  & 17.1  & 0.0036   & 0.0969    & 0.0005       \\
05:05:00            & GH            & -15.8           & 295.9     & 12.6      & +     & 0.0079  & 0.015        & 0.043  & 17  & 0.0042   & 0.0711    & 0.0003       \\
                          & GS            & -15           & 300     & 11.9      & +     & 0.0124  & 0.4758        & 0.1014  & 22.4  & 0.006   & 0.071    &         \\
\hline
Wind20140817    & L                & 73.1          & 290.1     & 9.3      & +     & 0.01  & 0.182        & 0.04  & 19  & 0.003   & 0.0873    & 0.0003       \\
07:40:00            & LA              & 67.3          & 316.4     & 8.9      & +     & 0.0096  & 0.063        & 0.038  & 27  & 0.0026   & 0.0773    & 0.00025       \\
10:22:00            & GH            & -71.3           & 123.2     & 9.7      & +     & 0.0099  & 0.112        & 0.034  & 27.1  & 0.0032   & 0.0861    & 0.00027       \\
                          & GS            & 67           & 320     & 7.6      & +     & 0.018  & 0.1167        & 0.107  & 13.7  & 0.007   & 0.084    &         \\
\hline
\end{tabular}
\end{footnotesize}
\\
  \small Note: Same parameters as in Table 1 but for {\it Wind}. 
\label{tab: fitting results - Wind}
\end{table}

%

%
\begin{acks}
  Support for this work came from the following grants: NASA STEREO Quadrature grant, NSF AGS-1435785, AGS-1433086 and AGS-1433213, and NASA NNX16AO04G, NNX15AB87G,  and NNX15AU01G. C. M. thanks the Austrian Science Fund (FWF): [P26174-N27].  P. V. is supported by an INSPIRE grant of AORC scheme under Department of Science and Technology. 
\end{acks}

Disclosure of Potential Conflicts of Interest

The authors declare that they have no conflicts of interest.

%
%
%

\end{article} 
\end{document}